\documentclass[aps,pra,nofootinbib,10pt,twocolumn]{revtex4-1}
\usepackage[colorlinks=true, citecolor=red, urlcolor=blue]{hyperref}
\usepackage[normalem]{ulem}
\usepackage[lmargin=2cm,rmargin=2cm,tmargin=2cm,bmargin=2cm]{geometry}
\usepackage{epsfig,newlfont,amssymb,amsfonts,amsmath,bm,subfigure,mathtools,amsthm,times,braket,soul}

\begin{document}
\title{Monogamy of quantum correlations - a review}
\author{Himadri Shekhar Dhar} 
\affiliation{Harish-Chandra Research Institute, Chhatnag Road, Jhunsi, Allahabad 211 019, India}
\affiliation{Homi Bhabha National Institute, Training School Complex, Anushaktinagar, Mumbai 400 085, India}
\author{Amit Kumar Pal}
\affiliation{Harish-Chandra Research Institute, Chhatnag Road, Jhunsi, Allahabad 211 019, India}
\affiliation{Homi Bhabha National Institute, Training School Complex, Anushaktinagar, Mumbai 400 085, India}
\author{Debraj Rakshit} 
\affiliation{Harish-Chandra Research Institute, Chhatnag Road, Jhunsi, Allahabad 211 019, India}
\affiliation{Homi Bhabha National Institute, Training School Complex, Anushaktinagar, Mumbai 400 085, India}
\author{Aditi Sen(De)}
\affiliation{Harish-Chandra Research Institute, Chhatnag Road, Jhunsi, Allahabad 211 019, India}
\affiliation{Homi Bhabha National Institute, Training School Complex, Anushaktinagar, Mumbai 400 085, India}
\author{Ujjwal Sen}
\affiliation{Harish-Chandra Research Institute, Chhatnag Road, Jhunsi, Allahabad 211 019, India}
\affiliation{Homi Bhabha National Institute, Training School Complex, Anushaktinagar, Mumbai 400 085, India}

\begin{abstract}
Monogamy is an intrinsic feature of quantum correlations that gives rise to several interesting quantum characteristics which are not amenable to classical explanations. The monogamy property imposes physical restrictions on unconditional sharability of quantum correlations between the different parts of a multipartite quantum system, and thus has a direct bearing on the cooperative properties of states of multiparty systems, including large many-body systems. On the contrary, a certain party can be maximally classical correlated with an arbitrary number of parties in a multiparty system. In recent years, the monogamy property of quantum correlations has been applied to understand several key aspects of quantum physics, including distribution of quantum resources, security in quantum communication, critical phenomena, and quantum biology. In this chapter, we look at some of the salient developments and applications in quantum physics that have been closely associated with the monogamy of quantum discord, and ``discord-like" quantum correlation measures. 
\end{abstract}

\maketitle 

\section{Introduction}
\label{sec:intro}

\begin{figure}
 \includegraphics[scale=0.25]{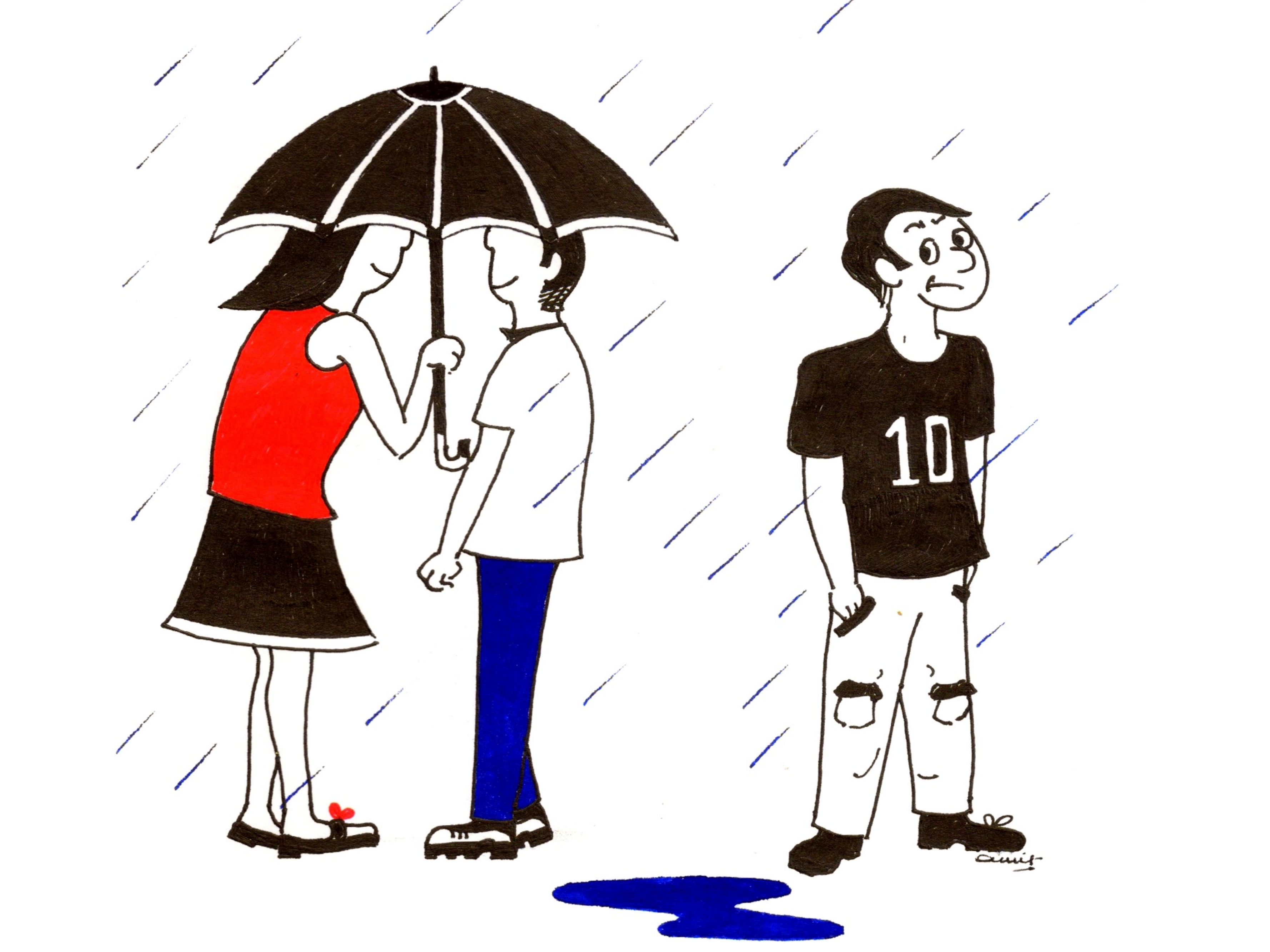}
 \caption{(Color online) Monogamy: Two persons sharing an umbrella are unmindful to the presence of the third person.}
 \label{monogamy-schematic}
\end{figure}

Quantum correlations, shared between two or more parties \cite{Horodecki09, Modi12}, boast of novel features, which are exclusive to the quantum world and are central to quantum information science. On one hand, they play a significant role in efficient quantum communication \cite{Gisin02,Terhal04,SenDe10} and computational \cite{Raussendorf01,Briegel09} tasks, while on the other hand, they help us to understand cooperative phenomena in quantum many-body systems \cite{Lewenstein07, Amico08, Augusiak12}.  A key difference between classical and quantum correlations is the way they can be shared among various parts of a multiparty quantum system. Unlike classical correlations, which can be freely shared, the sharability of quantum correlations is restricted by the non-classical properties of the quantum system. For example, in a tripartite quantum state, $\rho_{ABC}$, if two parties $A$ and $B$ are maximally quantum correlated, then none of $A$ and $B$ can share any quantum correlation with the third party, $C$ \cite{Ekert91,Bennett96,Terhal04,Coffman00,Kim12} (For a social representation, see Fig. \ref{monogamy-schematic}). However, there exists no such constraints for the classical correlations, and the pairs of parties, say $AB$ and $AC$, can simultaneously share maximum classical correlations. Similar situation arises in multiparty quantum systems of more than three parties. This exclusive trade-off between quantum correlations of different combinations of parties in a multiparty quantum system is known as {\it monogamy} of quantum correlations \cite{Terhal04,Ekert91,Bennett96,Coffman00,Kim12}. The no-go theorems, like the no-cloning theorem \cite{Wootters82,Dieks82,Yuen86,Barnum96}, put restrictions on the available options in quantum cryptography \cite{Gisin02}. Similarly, the monogamy of quantum correlations is a restriction on the sharability of quantum correlations, and yet help in obtaining advantages in a quantum system over their classical counterparts.  %In the same spirit as the beneficial role \cite{Gisin02,SenDe10} of several constraints \cite{Wootters82,Dieks82,Yuen86,Barnum96,Pati00,Kalev08} put on a single quantum system by quantum mechanical principles, the distribution of quantum correlations in the multipartite domain, often helps in obtaining quantum advantages over their classical counterparts in setting up quantum protocols in physical systems \cite{Giorgi11,Prabhu12,Kumar16,Allegra11,Song13,Qiu14,Qin16,Rao13,Zhu12,Chanda14}. 

In the seminal work by Coffman, Kundu, and Wootters (CKW) \cite{Coffman00}, a monogamy relation for three-qubit pure states was established by using an entanglement measure, namely, the squared concurrence \cite{Hill97,Wootters98}. This scenario was later generalized by Osborne and Verstraete \cite{Osborne06} for pure as well as mixed states of an arbitrary number of qubits. Since quantum correlations do not have a unique quantification even in the bipartite domain,  the immediate question that follows is whether the monogamy inequality proposed by CKW is necessarily obeyed by all kinds of quantum correlation measures. In general, quantum correlations can be broadly classified into two categories -- entanglement measures \cite{Horodecki09},  and the information-theoretic measures of quantum correlations \cite{Modi12}. While entanglement of formation \cite{Bennett96,Bennett96a,Heyden01}, concurrence \cite{Hill97,Wootters98}, distillable entanglement \cite{Rains99,Rains99a}, negativity \cite{Zyczkowski98,Peres96,Horodecki96,Lee00,Vidal02,Plenio05}, logarithmic negativity \cite{Zyczkowski98,Lee00,Vidal02,Plenio05}, relative entropy of entanglement \cite{Vedral97,Vedral98,Vedral02}, etc. belong to the first category, quantum discord \cite{Henderson01,Ollivier02}, and quantum work deficit \cite{Oppenheim02,Horodecki03,Devetak05,Horodecki05} are examples of the second kind. Although quantum correlations are qualitatively monogamous, not all of them are limited by only the form of monogamy constraint proposed by CKW. In particular, while the squared concurrence and negativity satisfy the CKW monogamy inequality  for all three-qubit pure states \cite{Coffman00,Osborne06,Ou07}, many others, such as logarithmic negativity and  information-theoretic measures, do not satisfy the same \cite{Ou07,Giorgi11,Prabhu12,Fanchini11,Streltsov12}. Deliberations on the monogamy of quantum correlations have led to important insights including a ``conservation law" between entanglement and information-theoretic quantum correlations in multiparty quantum states \cite{Fanchini11}. In Ref. \cite{Lancien16}, the authors formulate the requirements for a bipartite entanglement measure to be monogamous for all quantum states, and show that additive and suitably normalized entanglement measure, which can faithfully describe the geometric structure of the fully antisymmetric state, are non-monogamous. However, it is also understood that all kinds of quantum correlations obey the CKW monogamy constraint for a given state when raised to a suitable power, provided they follow certain conditions \cite{Salini14}.

Interestingly, the limitation imposed by the quantum mechanical principles, in form of monogamy constraints, is not exclusive to quantum correlations. There exists no-go theorems which place parallel restrictions such as monogamy of Bell inequality violation \cite{Toner09,Kurzynski11} and exclusion principle of classical information transmission over quantum channels \cite{Prabhu13} (cf. \cite{Oliveira12,Sun13,Sadhukhan15,Prabhu13a}). More precisely,  within the consideration of a multiparty set-up, for example, of an editor with several reporters, if the shared quantum state between the editor and a single reporter violates a Bell inequality \cite{Bell64,Clauser69} or is quantum dense codeable \cite{Bennett92}, then the rest of the channels shared between the editor and the other reporters are prohibited from possessing the same quantum advantage. Analogous monogamy constraints have also been addressed in the context of quantum steering \cite{Reid13,Milne14}, quantum teleportation fidelity \cite{Lee09}, and contextual inequalities \cite{Ramanathan12,Kurzynski14}.    

The monogamy properties of quantum correlations find potential applications in quantum information based protocols like quantum cryptography  \cite{Gisin02}, entanglement distillation \cite{Bennett96a}, quantum state and channel discrimination \cite{Giorgi11,Prabhu12,Kumar16}, and in characterizing quantum many-body systems \cite{Allegra11,Song13,Qiu14,Qin16,Rao13,Chandran07,Dhar11,Roy16,Sadhukhan16} as well as in biological processes \cite{,Zhu12,Chanda14}. The key concept of entanglement-based quantum cryptography essentially exploits the trade off in monogamy of quantum correlations, which limits the amount of information that an eavesdropper can extract about the secret key, shared between a sender and a receiver, obtained via measurement on both sides of an entangled state between the sender and the receiver \cite{Terhal04,Ekert91,Bera16}. The constraints on shareability of entanglement finds further application in enhancing quantum privacy via entanglement purification \cite{Deutsch96}. Another importance of  monogamy relations, arising due to the constraints that they put on the distribution of quantum correlation among many parties, is their ability to capture multipartite quantum correlations present in the system, the latter being, in general,  a challenging task \cite{Coffman00,Bera12,Prabhu12a}. Moreover, they play a decisive role in designing the structure of eigenstates of quantum spin models, which are expected to obey the no-go principles arising from the monogamy constraints \cite{Sadhukhan16}. 
 
In this chapter, our main aim is to review the results on monogamy of quantum discord, and other ``discord-like" measures of quantum correlations. We survey the properties of a proposed multiparty quantum correlation measure, called the ``monogamy score", and its relations with other measures of quantum correlations. Finally, we also take a look at the usefulness of the monogamy score for quantum discord in quantum information science, and in quantum many-body physics. The structure of this chapter is as follows. In Section \ref{sec:ent-mono}, we present a short review on the studies that have been carried out on the monogamy properties of entanglement. Section \ref{sec:measures} consists of the definitions of several quantum correlation measures, such as quantum discord, quantum work deficit, and geometric quantum discord, that have been defined independent of the entanglement-separability paradigm, The monogamy properties of these measures are discussed in the following sections. In Section \ref{sec:disc_mono}, we focus on the monogamy of quantum discord. The monogamy property of other information-theoretic quantum correlation measures are discussed in Section \ref{sec:mon-other}. Section \ref{sec:oth-meas} provides a report on the relation between the monogamy of quantum correlations with other multiparty measures. In Section \ref{sec:mono-appli}, we review some noteworthy applications of the monogamy property. Section \ref{conclusion} contains some  concluding remarks.

\section{Monogamy relations of entanglement} 
\label{sec:ent-mono}

In this section, we provide a brief discussion on the monogamy properties of different entanglement measures, such as concurrence, negativity, etc.  Starting from tripartite quantum systems, we expand the discussion to the more complex cases in multiparty systems. For a review, see \cite{Kim12}.

We also formally define the ``monogamy score" corresponding to an arbitrary quantum correlation measure. 
  
\subsection{Tripartite system: CKW inequality and beyond}  
\label{subsec:three-party-entanglement}

In this subsection, we review monogamy properties of entanglement in the tripartite scenario, using concurrence as the entanglement measure. For an arbitrary two-qubit quantum state $\rho_{AB}$, concurrence is defined as \cite{Hill97, Wootters98} $\mathcal{C}_{AB}\equiv\mathcal{C}(\rho_{AB})= \max\{0,\lambda_1-\lambda_2-\lambda_3-\lambda_4\}$, where $\{\lambda_i\},\,,i=1,\ldots,4,$ are square roots of the eigenvalues of the positive operator $\rho\tilde{\rho}$ in descending order, and the spin-flipped density matrix $\tilde{\rho}$ is given by $\tilde{\rho} = (\sigma_y \otimes \sigma_y) \rho^* (\sigma_y \otimes \sigma_y)$. Here and henceforth, $\sigma_x$, $\sigma_y$ and $\sigma_z$ denote the standard Pauli spin matrices.  Note that $\mathcal{C}$ vanishes for the separable states and attains the value $1$ for the maximally entangled states in $\mathbb{C}^2 \otimes \mathbb{C}^2$ systems. The physical significance of concurrence stems from the fact that the entanglement of formation is a monotonic function of concurrence, and vice versa,  in $\mathbb{C}^2 \otimes \mathbb{C}^2$ \cite{Hill97, Wootters98}. A formal definition of entanglement of formation is given later in this section.

Consider a three-qubit pure state, $\rho_{ABC}=(|\phi\rangle\langle \phi|)_{ABC}$. The concurrences corresponding to the the reduced density matrices  $\rho_{AB}=\text{Tr}_C(\rho_{ABC})$ and $\rho_{AC}=\text{Tr}_B(\rho_{ABC})$ satisfy the inequalities, $\mathcal{C}_{AB}^2 \le \text{Tr}(\rho_{AB} \tilde{\rho}_{AB})$ and $\mathcal{C}_{AC}^2 \le \text{Tr}(\rho_{AC} \tilde{\rho}_{AC})$, respectively.  It can further be shown that $\text{Tr}(\rho_{AB} \tilde{\rho}_{AB})+ \text{Tr}(\rho_{AC} \tilde{\rho}_{AC}) = 4~\text{det} \rho_A$. This leads to following inequality \cite{Coffman00}:
\begin{eqnarray}
\label{eq1}
\mathcal{C}_{AB}^2+\mathcal{C}_{AC}^2 \le 4 \det \rho_A,
\end{eqnarray}
where $\rho_A=\text{Tr}_{BC}(\rho_{ABC})$.  Even though $BC$ is a four-dimensional system, the support of $\rho_{BC}=\text{Tr}_{A}(\rho_{ABC})$ is spanned by the eigenstates corresponding to at most two non-zero eigenvalues of the reduced density matrix $\rho_{BC}$, and hence is effectively defined on a two-dimensional space. This allows to treat the bipartite split of $A$ and $BC$ as an effective two-qubit system whose concurrence, $\mathcal{C}_{A:BC}$, is simply $2\sqrt{\det \rho_A}$. As a result, the inequality in  (\ref{eq1}) becomes
\begin{eqnarray}
\label{eq2}
\mathcal{C}_{AB}^2+\mathcal{C}_{AC}^2 \le \mathcal{C}_{A:BC}^2,
\end{eqnarray}
and is referred to as the monogamy inequality for squared concurrence in the case of three-qubit pure state. For a three-qubit mixed state, the state $\rho_{BC}$ may, in principle, have all four non-zero  eigenvalues. However, a generalization of the above inequality for the mixed state is prescribed by replacing the right hand side of (\ref{eq2}) by the minimum average concurrence squared over all possible pure state decompositions \{$p_i,\ket{\psi_i}$\} of $\rho_{BC}$, and hence  (\ref{eq2}) also holds for arbitrary mixed three-qubit states.  

In this context, let us define the tangle for a three-qubit pure state, $\rho_{ABC}$,  expressed in terms of squared concurrences, as \cite{Coffman00}
\begin{eqnarray}
\label{eq3}
\tau_{ABC} = \mathcal{C}_{A:BC}^2-\mathcal{C}_{AB}^2-\mathcal{C}_{AC}^2.
\end{eqnarray}
It turns out that $\tau_{ABC}$, which is also known as the \emph{three-tangle}, or the \emph{residual entanglement}, is independent of the choice of the ``node", which is the site $A$ here. The tangle has been argued to  characterize three-qubit entanglement. The generalization of the tangle to mixed states can be obtained by the convex roof extension, the computation of which is difficult. 

Next, let us present the definition of the  entanglement of formation (EoF) \cite{Wootters98}, a measure of bipartite entanglement, and its relation with concurrence for two-qubit states. Consider a bipartite quantum state $\rho_{AB}$, and the  ensemble $\{p_i,|\psi_i\rangle\}$ denoting a possible pure state decomposition of $\rho_{AB}$, satisfying $\rho_{AB}=\sum_{i}p_i|\psi_i\rangle\langle\psi_i|$. The EoF is defined as 
\begin{eqnarray}
\label{eof}
E_f(\rho_{AB})=\underset{\{p_i,|\psi_i\rangle\}}{\min} \sum_{i} p_i S(\text{Tr}_B[|\psi_i\rangle\langle\psi_i|]),
\end{eqnarray}
where $S(\text{Tr}_B[|\psi_i\rangle\langle\psi_i|])$ is the von Neumann entropy of the reduced density matrix corresponding to the $A$ part of $\rho_{AB}$. A compact formula for the EoF is known for two-qubit systems in terms of the concurrence, $\mathcal{C}_{AB}$. For a two-qubit mixed state $\rho_{AB}$, $E_f(\rho_{AB})=h \left((1+\sqrt{1-\mathcal{C}_{AB}^2})/2\right)$ \cite{Wootters98}, where $h(x)=-x \log_2 x-(1-x)\log_2(1-x)$. The EoF, being a concave function of squared concurrence, does not obey the CKW inequality. However, the EoF can also not be freely shared amongst the constituents of a multiparty system. In fact, the square of the EoF does obey the same relation as the squared concurrence for tripartite systems \cite{Bai14}.

Several studies have addressed the monogamy property of concurrence within the tripartite scenario. We briefly mention some important findings in this direction. It is natural to ask whether the monogamy inequality is satisfied in tripartite systems consisting of higher-dimensional parties. The analysis understandably becomes complex with increasing dimension of the Hilbert space, as much less is known about the quantification of entanglement in higher dimensional cases. For example, the exact formula for EoF and concurrence are missing in higher dimensions.  However, there have been definitive efforts to understand the monogamy constraints in higher dimensional systems. The results reported in Ref.~\cite{Ou07a} indicate that the CKW inequality cannot be directly extended to higher-dimensional states.  However, it has been demonstrated that the squared concurrence satisfies the monogamy relation for arbitrary pure states in $\mathbb{C}^2 \otimes \mathbb{C}^2 \otimes \mathbb{C}^4$    \cite{Ren10} (cf. \cite{Yu08,Huang08}). The monogamy property of squared concurrence in higher-dimensional systems of more than three parties has also been addressed \cite{Ou08}. Other approaches to construct more monogamy inequalities for entanglement in tripartite states have made use of the generalized concurrence \cite{Cornelio13}, which is a multipartite measure of entanglement. 

\subsection{Monogamy score}
\label{subsec:score}

Just like for squared concurrence, one can formulate the problem of monogamy for arbitrary quantum correlation measures. For a given bipartite quantum correlation measure, $Q$, and a three-party quantum state, $\rho_{ABC}$, we call the state to be monogamous for the measure $Q$ if 
\begin{eqnarray}
Q(\rho_{A:BC})\geq Q(\rho_{AB})+Q(\rho_{AC}).
\label{eq:monogamy}
\end{eqnarray}
Otherwise, the state is non-monogamous for that measure. The measure is called monogamous for a given tripartite quantum system if it is monogamous for all tripartite states. Similar to the spirit of tangle defined in Eq. (\ref{eq3}), one can define a quantity called \emph{monogamy score} \cite{Bera12}, of a bipartite quantum correlation measure, $Q$, based on the monogamy relation, given in  (\ref{eq:monogamy}). It is given by 
\begin{eqnarray}
\delta_Q=Q(\rho_{ABC})- Q(\rho_{AB})-Q(\rho_{AC}),
\label{eq:score}
\end{eqnarray}
where we call the party ``$A$" as the nodal observer. With this notion, the tangle \cite{Coffman00} can be called monogamy score for squared concurrence, and can also alternatively be denoted by $\delta_{\mathcal{C}^2}$.  This definition can be extended to an $N$-party quantum state, $\rho_{12\ldots N}$. An arbitrary $N$-party state is said to be monogamous with respect to $Q$, if 
\begin{eqnarray}
Q(\rho_{j:rest}) \geq \sum_{k\neq j} Q(\rho_{jk}),
\label{eq:monogamy_nparty}
\end{eqnarray}
and the corresponding monogamy score, for any quantum correlation measure, $Q$, is defined as 
\begin{equation}
\delta^{j}_Q(\rho_{1:23\ldots N}) = Q(\rho_{j:rest}) - \sum_{k\neq j} Q(\rho_{jk}),
\label{score_nparty}
\end{equation}
with $j$ as the node. Here, $\rho_{jk}$ represents the two-party density matrix, which can be obtained from $\rho_{12\ldots N}$ by tracing out  all the other parties except $j$ and $k$ ($j,k=1,2,\ldots,N$).  It has been argued \cite{Coffman00,Bera12} that the monogamy score quantifies a multiparty quantum correlation measure, and hence constitutes a method of conceptualizing a multiparty measure using bipartite ones. A monogamous quantum correlation measure, for a given node $j$, have $\delta_Q^{j}\geq 0$ for all multipartite states. Unless otherwise stated, we always use first party as the nodal observer and in that case, we denote monogamy score as \(\delta_Q\), discarding the superscript.  Henceforth, we shall always describe an $N$-party quantum state, pure or mixed, by $\rho_{12\ldots N}$, while for ease of notations, we denote a three-party quantum state by $\rho_{ABC}$. 

%\subsection{Multiparty systems: Generalization of CKW inequality}  
%\label{subsec:ckw-multiparty}

In an $N$-party scenario, Coffman, Kundu, and Wootters conjectured a generalization of the CKW inequality to $N$-qubit pure states, $\rho_{12\ldots N}$, i.e., the inequality in (\ref{eq:monogamy_nparty}), by replacing $Q$ with $\mathcal{C}^2$. Some time later, the conjecture was proven for arbitrary $N$-qubit states by Osborne and Verstraete \cite{Osborne06}, by using an inductive strategy. Also, there have been several attempts to construct generalized monogamy inequalities for entanglement in qubit systems \cite{Eltschka09,Eltschka14,Gour10,Cornelio13,Regula14,Zhu14}. 
%Ref.~\cite{Regula14}  has proposed a set of monogamy inequalities that sharpens the conventional CKW constraints. Finally, we would like to mention about the generalized monogamy inequalities given by the $\alpha${th} power of concurrence and EoF for $N$-qubit states \cite{Zhu14}, based on which the tangle and EoF have been studied. 

\subsection{Monogamy of negativity and other entanglement measures} 
\label{subsec:neg}

The negativity \cite{Peres96,Horodecki96,Zyczkowski98,Lee00,Vidal02,Plenio05}, $\mathcal{N}(\rho_{AB})$, corresponding to the quantum state $\rho_{AB}$ defined on the Hilbert space  $\mathbb{C}_A \otimes \mathbb{C}_B$ for two parties $A$ and $B$, is defined by $\mathcal{N}_{AB}\equiv\mathcal{N}(\rho_{AB})=(\|\rho_{AB}^{T_A}\|_1-1)/2$, where $\rho_{AB}^{T_A}$ is obtained by performing the partial transposition on the state $\rho_{AB}$ with respect to the subsystem $A$ \cite{Peres96,Horodecki96}, i.e., $(\rho^{T_A})_{ij,kl}=(\rho)_{kj,il}$, and where $\|\rho\|_1 = \mbox{Tr}\sqrt{\rho \rho^{\dagger}}$ denotes the trace norm of the matrix $\rho$. For systems in $\mathbb{C}^2 \otimes \mathbb{C}^2$ and $\mathbb{C}^2 \otimes \mathbb{C}^3$ systems, $\mathcal{N} > 0$ is the necessary and sufficient for non-separability. In order to achieve a maximum value of unity in $\mathbb{C}^2 \otimes \mathbb{C}^2$, the negativity can be redefined as $\mathcal{N}_{AB}=\|\rho_{AB}^{T_A}\|_1-1$.

For any pure three-qubit state $\rho_{ABC}$, the monogamy inequality for squared negativity \cite{Ou07},
\begin{eqnarray}
\label{eq7}
\mathcal{N}_{AB}^2+\mathcal{N}_{AC}^2 \le \mathcal{N}_{A:BC}^2,
\end{eqnarray}
holds, where $A$ has been chosen as the nodal party, and  in  (\ref{eq:monogamy}), $Q$ is replaced by $\mathcal{N}^2$. For any three-qubit pure state, it turns out that $\mathcal{N}_{A:BC}=\mathcal{C}_{A:BC}$. Moreover, as shown in Ref.~\cite{Chen05} for arbitrary two-qubit mixed states, we have  
\begin{eqnarray}
\label{eq8}
\mathcal{N}_{AB}=\|\rho_{AB}^{T_A}\|_1-1 \le \mathcal{C}_{AB}. 
\end{eqnarray}
Thus for any three-qubit pure state, we obtain $\mathcal{N}_{AB} \le \mathcal{C}_{AB}$ and $\mathcal{N}_{AC} \le \mathcal{C}_{AC}$, and consequently, the proof of (\ref{eq7}) automatically follows from the corresponding one for squared concurrence. For an $N$-qubit pure state, the generalized monogamy inequality for squared negativity, given in Eq. (\ref{eq:monogamy_nparty}) can be proven by using $\mathcal{N}_{1:23 \ldots N}=\mathcal{C}_{1:23 \ldots N}$ for pure states, and the relation in Eq. (\ref{eq8}).  

As mentioned in the previous subsection, there have been efforts to propose stronger versions of the monogamy relation for entanglement. 
Recently, a stronger monogamy inequality for negativity has been proposed \cite{Karmakar16}, and a detailed study has been performed for four-qubit states. Another interesting proposal pitches the square of convex-roof extended negativity as an alternative candidate to characterize strong monogamy of multiparty quantum entanglement \cite{Choi15}. 

Monogamy has also been studied for other entanglement measures, including entanglement of assistance (EoA) \cite{DiVincenzo98},  and squashed entanglement \cite{Koashi04, Christandl04}. The definition of EoA, which was originally introduced in terms of entropy of entanglement,  can, in principle, be generalized for other measures of entanglement. For example, concurrence of assistance (CoA) is an entanglement monotone for pure tripartite states \cite{Gour05}, and similar to the squared concurrence, monogamy properties of the squared CoA have been studied extensively \cite{Gour05,Zhu15,Oliveira14,Bai14}.  Ref.~\cite{Li09} found lower and upper bounds of EoA, among which the upper bound was shown to obey monogamy constraints for arbitrary $N$-qubit states. In Ref. \cite{Koashi04}, Koashi and Winter showed that for arbitrary tripartite states, the one-way distillable entanglement, the one-way distillable secret key \cite{Devetak03}, and the squashed entanglement \cite{Christandl04} satisfy the monogamy relation, given in  (\ref{eq:monogamy}). For tripartite pure states, the entanglement of purification \cite{Terhal02} was shown to be non-monogamous in general \cite{Bagchi15}.

The monogamy inequality of entanglement sharing has been investigated also for continuous variable systems.  In Refs.~\cite{Addesso06,Addesso06a}, monogamy inequality for the ``continuous variable tangle",  or the ``contangle", has been provided for arbitrary three-mode Gaussian states and for symmetric arbitrary-mode Gaussian states, where contangle is defined as the convex roof of the square of the logarithmic negativity.  Further generalization of the results have been achieved in  \cite{Addesso07,Hiroshima07,Addesso08}.

\section{Information-Theoretic Measures of Quantum Correlation}
\label{sec:measures}

In this section, we define a few information-theoretic measures of quantum correlations, whose monogamy properties are discussed in the subsequent sections. 

\subsection{Quantum discord}
\label{subsec:qd}
Let us consider a bipartite quantum state $\rho_{AB}$, for which the \textit{uninterrogated}, or \textit{unmeasured} quantum conditional entropy is defined as 
\begin{eqnarray}
 \tilde{S}(\rho_{A|B}) = S(\rho_{AB}) - S(\rho_B),
 \label{q-cond-ent-unmeas}
\end{eqnarray}
where $\rho_{B}=\mbox{Tr}_{A}(\rho_{AB})$ is the reduced density matrix of the subsystem $B$, obtained by tracing over the subsystem $A$. 
One can also define an \textit{interrogated} conditional entropy,  given by 
\begin{eqnarray}
S(\rho_{A|B}) = \min_{\{\Pi_i^B\}} \sum_i p_i S(\rho_{A|i}),
\end{eqnarray}
where the minimization is performed over all complete sets of projective measurements, \(\{\Pi^B_i\}\),  performed on subsystem \(B\). The corresponding post-measurement state  for subsystem \(A\) is given by \(\rho_{A|i} = \mbox{Tr}_B[(\mathbb{I}_A \otimes \Pi^B_i) \rho_{AB} (\mathbb{I}_A \otimes \Pi^B_i)]/p_i\), where \(\mathbb{I}_A\) is the identity operator on the Hilbert space of the subsystem \(A\), and 
\(p_i = \mbox{Tr}_{AB}[(\mathbb{I}_A \otimes \Pi^B_i) \rho_{AB} (\mathbb{I}_A \otimes \Pi^B_i)]\) is the probability of obtaining the outcome \(i\). These two definitions of quantum conditional entropy are two different extensions of the two equivalent expressions of the classical mutual information. The former is used to define the uninterrogated quantum mutual information, given by 
\begin{eqnarray}
 \tilde{I}(\rho_{AB})=S(\rho_A) - \tilde{S}(\rho_{A|B}),
\end{eqnarray}
which is interpreted as the ``total correlation'' of $\rho_{AB}$\cite{Zurek83,Barnett89,Schumacher96,Cerf97,Groisman05,Ollivier02,Henderson01}. On the other hand,  the latter provides the definition of the interrogated quantum mutual information, 
\begin{eqnarray}
 I^{\leftarrow}(\rho_{AB})=S(\rho_A) - S(\rho_{A|B}),
 \label{class-corr}
\end{eqnarray}
also interpreted as the ``classical correlation'' present in the quantum state $\rho_{AB}$ \cite{Ollivier02,Henderson01}. The arrow in the superscript begins from the subsystem on which the measurement is performed. The quantum discord of the state $\rho_{AB}$ is the difference between the uninterrogated and interrogated quantum mutual informations \cite{Ollivier02,Henderson01}, and is given by 
\begin{eqnarray}
 D^{\leftarrow}(\rho_{AB})&=&\tilde{I}(\rho_{AB})-I^{\leftarrow}(\rho_{AB})\nonumber \\ &=& S(\rho_{A|B})-\tilde{S}(\rho_{A|B}).
 \label{eq:qd}
\end{eqnarray}

Note here that one can also define quantum discord, $D^{\rightarrow}(\rho_{AB})$, by performing local measurement over the subsystem $A$ 
instead of the subsystem $B$. In general, $D^{\leftarrow}(\rho_{AB})\neq D^{\rightarrow}(\rho_{AB})$. Unless otherwise stated, here and throughout in this chapter, we consider $D^{\leftarrow}(\rho_{AB})$ as the measure of quantum discord. Note also that the definition of quantum discord is provided by using local projective measurement. However, quantum discord can also be defined in terms of positive operator valued measurements (POVMs). 

\subsection{Quantum work deficit}
\label{subsec:qwd}

The amount of extractable pure states from a bipartite state $\rho_{AB}$, under a set of global operations, called the  \textquotedblleft closed operations\textquotedblright (CO), is given by \cite{Oppenheim02,Horodecki03,Devetak05,Horodecki05}
\begin{eqnarray}
 I_{\mbox{\scriptsize CO\normalsize}}=\log_{2}\mbox{dim}\left(\mathcal{H}\right)-S(\rho_{AB}),
\end{eqnarray}
where the set of closed operations consists of \textit{(i)} unitary operations, and \textit{(ii)} dephasing the bipartite state by a set of projectors, $\{\Pi_{k}\}$, defined  on the Hilbert space $\mathcal{H}$ of $\rho_{AB}$. On the other hand, considering the set of `` closed local operations and classical communication" (CLOCC), the amount of extractable pure states from $\rho_{AB}$ is given by \cite{Oppenheim02,Horodecki03,Devetak05,Horodecki05}
\begin{eqnarray}
 I_{\mbox{\scriptsize CLOCC\normalsize}}=\log_{2}\mbox{dim}\left(\mathcal{H}\right)-\min
 S\left(\rho_{AB}^{\prime}\right).
\end{eqnarray}
Here, CLOCC consists of \textit{(i)} local unitary operations, \textit{(ii)} dephasing by local measurement on the subsystem, say, $B$, and \textit{(iii)} communicating the dephased subsystem to the other party, $A$, via a noiseless quantum channel. The average quantum state, after the local projective measurement $\{\Pi_{k}^{B}\}$ is performed on $B$, can be written as  $\rho_{AB}^{\prime}=\sum_{k} p_{k}\rho_{AB}^{k}$ with $\rho_{AB}^{k}$ and $p_{k}$ being defined in a similar fashion as in the case of quantum discord. The minimization in $I_{\mbox{\scriptsize CLOCC\normalsize}}$ is achieved over all complete sets $\{\Pi_{k}^{B}\}$. The ``one-way'' quantum work deficit is then defined as \cite{Oppenheim02,Horodecki03,Devetak05,Horodecki05}
\begin{eqnarray}
W^{\leftarrow}(\rho_{AB})&=&I_{\mbox{\scriptsize CO\normalsize}}-I_{\mbox{\scriptsize CLOCC\normalsize}}\nonumber\\
&=&\underset{\{\Pi_k^B\}}{\min}\left[ S\left(\rho_{AB}^{\prime}\right)-S(\rho_{AB})\right],
\end{eqnarray} 
where similar to quantum discord, the the arrow in the superscript starts from the subsystem over which the measurement is performed. 

\subsection{Geometric measure of quantum discord}
\label{subsec:gqd}

Geometric quantum discord \cite{Dakic10,Luo10} for a bipartite quantum state $\rho_{AB}$ can be defined as the minimum squared Hilbert-Schmidt distance of $\rho_{AB}$ from the set, $\mathcal{S}_{QC}$, of all ``quantum-classical'' states, given by $\sigma_{AB}=\sum_{i}p_i\sigma_A^i\otimes\ket{i}\bra{i}$, where $\{\ket{i}\}$ forms a mutually orthonormal set of the Hilbert space of the subsystem $B$. Mathematically, geometric  quantum discord is given by   
\begin{eqnarray}
 D_G(\rho_{AB})=\underset{\sigma_{AB}\in \mathcal{S}_{QC}}{\min}||\rho_{AB}-\sigma_{AB}||^2_2,
\end{eqnarray}
where  $||\rho-\sigma||_2=\mbox{Tr}(\rho-\sigma)^{2}$, for two arbitrary density matrices $\rho$ and $\sigma$ (however, see \cite{Piani12}). Although geometric measures can be defined by using general Schatten $p$-norms \cite{Debarba12} (see also \cite{Rana13,Debarba13}), it has been shown in Ref. \cite{Paula13} that the geometric quantum discord can be consistently defined by using the one-norm, i.e., the trace-distance only. Note here that the asymmetry in the definition of quantum discord due to a local measurement over one of the parties also remains here in the choice of $\sigma_{AB}$. One can also consider ``classical-quantum'' states, $\tilde{\sigma}_{AB}=\sum_jp_j\ket{j}\bra{j}\otimes\sigma_B^j$, to define the geometric quantum discord.

\section{Monogamy of Quantum Discord}
\label{sec:disc_mono}

The monogamy score for quantum discord,  defined after (\ref{eq:score}), is denoted by $\delta_{D}^\leftarrow$ or $\delta_{D}^\rightarrow$, depending on the subsystem over which the measurement is performed while computing the quantum discord. In Refs. \cite{Prabhu12,Giorgi11}, it was found that quantum discord violates the monogamy relation already for certain three-qubit pure states.

Before discussing the monogamy properties of quantum discord in detail, let us first ask the question as to whether a measure of  quantum correlation, chosen from the information-theoretic domain, can be monogamous. Although this is a difficult question to answer in its full generality, some insight can be obtained by noting that there is a marked difference between such a measure and the ones belonging to the entanglement-separability category. The former may have a non-zero value in the case of a separable state, while by definition, entanglement measures vanish for all unentangled states. Let us consider a general bipartite quantum correlation measure, $Q$, which, for an arbitrary bipartite quantum state $\rho_{AB}$, obeys a set of basic properties, as enumerated below \cite{Streltsov12}. 
\begin{enumerate}
 \item[\textbf{P1.}] Positivity: $Q(\rho_{AB})\geq0$.
 \item[\textbf{P2.}] Invariance under local unitary transformation: $Q(\rho^\prime_{AB})=Q(\rho_{AB})$, with  $\rho^\prime_{AB}=(U_A\otimes U_B)\rho_{AB}(U_A^\dagger\otimes U_B^\dagger)$. Here, $U_A$ and $U_B$ are unitary operators defined on the Hilbert spaces of the subsystems $A$ and $B$. 
 \item[\textbf{P3.}] Non-increasing upon the introduction of a local pure ancilla: $Q(\rho_{AB})\geq Q(\tilde{\rho}_{A:BC})$, with 
 $\tilde{\rho}_{A:BC}=\rho_{AB}\otimes(\ket{0}\bra{0})_C$. 
\end{enumerate}
\noindent Note here that the first two properties are standard requirements for any measure of quantum correlations, i.e., both entanglement and information-theoretic measures. The third property is satisfied, for example, by quantum discord, irrespective of whether the ancilla is attached to the measured, or the unmeasured side. The direction of the inequality in \textbf{P3} is interesting. It may seem that we should have $Q(\rho_{AB})\leq Q_{A:BC}(\rho_{AB}\otimes(\ket{0}\bra{0})_C)$, as throwing out the $C$-part, of a state in $A:BC$, may only ``harm" (i.e., reduce $Q$), if at all. However, if we look at the definition of quantum discord, we find that \emph{having} the extra $C$-part may do harm, as it is only the $I^\leftarrow$ term in $D^\leftarrow$ that can get affected due to the extra $C$-part. The extra $C$-part leads to a larger class of possible measurements that can be performed for the maximization in $I^\leftarrow_{A:BC}$, than in $I^\leftarrow_{AB}$. 

A generic separable state of the $AC$ system is given by 
\begin{eqnarray}
\rho_{AC}=\sum_i p_iP_A(\ket{\psi_i})\otimes P_C(\ket{\phi_i}),
\label{sep-state}
\end{eqnarray}
where $P(\ket{\alpha})=\ket{\alpha}\bra{\alpha}$. Let us now consider  a special form of the tripartite separable state $\rho_{ABC}=\sum_i p_i P_A(\ket{\psi_i})\otimes P_B(\ket{i})\otimes P_C(\ket{\phi_i})$ with $\{\ket{i}\}$ being a set of mutually orthonormal states. The quantum correlation, $Q$, in the $A:BC$ bipartition, has the same value as that in the unitarily connected state $\sigma_{ABC}=\sum_i p_iP_A(\ket{\psi_i})\otimes P_B(\ket{i})\otimes P_C(\ket{0})$. Also, the amount of quantum correlation present in the state $\sigma_{ABC}$ in the $A:BC$ bipartition can not be higher than that present in the $A:B$ bipartition in the state $\sigma_{AB}=\mbox{Tr}_C(\sigma_{ABC})$, i.e., 
$Q(\sigma_{AB})\geq Q(\rho_{A:BC})$. If we now assume that  $Q$ satisfy the monogamy relation, then $Q(\sigma_{AB})\geq Q(\rho_{AB})+Q(\rho_{AC})$. Since $\rho_{AB}\equiv\sigma_{AB}$, we obtain $Q(\rho_{AC})\leq0$, which, due to  the positivity of $Q$, implies $Q(\rho_{AC})=0$. Hence, a quantum correlation measure, $Q$, which is monogamous, and which obeys the properties \textbf{P1}--\textbf{P3}, 
must be zero for a generic separable state, $\rho_{AC}$. Contrapositively, a general bipartite quantum correlation measure, $Q$, which has a non-zero value for at least one separable state, and which obeys a set of basic properties must be non-monogamous \cite{Streltsov12}.

\begin{figure}
 \includegraphics[scale=0.6]{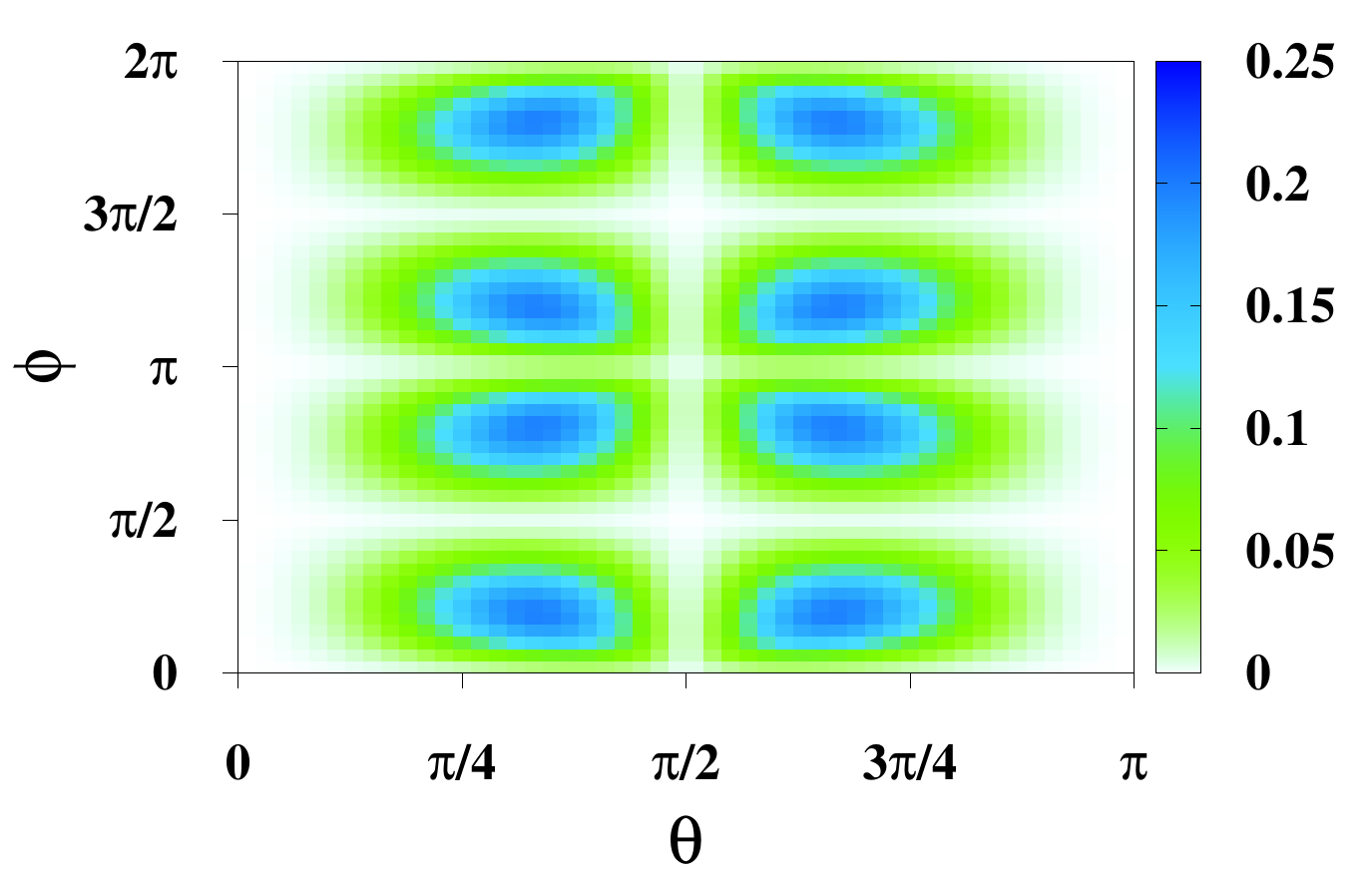}
 \caption{(Color online) Plot of $-\delta_D^\leftarrow$ as a function of $\theta$ and $\phi$ for three-qubit generalized W states given in Eq. (\ref{genw}). In agreement with the results  reported in \cite{Giorgi11,Prabhu12}, $-\delta_D^\leftarrow$ is positive over the entire plane of $(\theta,\phi)$. Reproduced figure with permission from the Authors and the Publisher of Ref. \cite{Prabhu12}. Copyright (2012) of the American Physical Society.}
 \label{mono_gw}
\end{figure}

The above discussion indicates that in the three-qubit scenario, monogamy for a general measure of quantum correlation, which belongs to the information-theoretic domain and which can be non-zero for a two-qubit separable state of rank $2$, can be violated for pure three-qubit states. This includes quantum discord, and other ``discord-like'' measures \cite{Modi12}. As an example, let us consider the three-qubit generalized W states \cite{Zeilinger92,Sen03}, parametrized using two real parameters and given by 
\begin{eqnarray}
\ket{\text{gW}}=\sin\theta\cos\phi\ket{100}+\sin\theta\sin\phi\ket{010}+\cos\theta\ket{001},\nonumber \\
\label{genw}
\end{eqnarray}
with $\theta,\phi$ ($0\leq\theta\leq\pi$, $0\leq\phi<2\pi$) being the real parameters. The plot of the negative of the quantity $\delta^\leftarrow_D=D^\leftarrow(\rho_{A:BC})-D^\leftarrow(\rho_{AB})-D^\leftarrow(\rho_{AC})$ as a function of $\theta$ and $\phi$ is presented in Fig. \ref{mono_gw}. Note that over the entire plane of $(\theta,\phi)$, $-\delta_D^\leftarrow$ has  positive values, indicating that generalized W states always violate monogamy of quantum discord \cite{Prabhu12,Giorgi11}. 

In this context, one must note that the monogamy inequality of quantum discord, as given by $D^{\rightarrow}(\rho_{A:BC})\geq D^{\rightarrow}(\rho_{AB})+D^{\rightarrow}(\rho_{AC})$, when the measurement is always performed over the nodal observer, can be shown to be equivalent to an inequality between the uninterrogated mutual information of the subsystem $BC$, and the EoF of the same, as given by \cite{Ren13}
\begin{eqnarray}
 E_f(\rho_{BC})\leq\frac{\tilde{I}(\rho_{BC})}{2},
 \label{ren-eq}
\end{eqnarray}
for three-qubit pure states. Li and Luo \cite{Li07} have shown that the inequality given in (\ref{ren-eq}) does not hold for all tripartite pure states, thereby indicating that quantum discord can be both monogamous as well as non-monogamous, complementing the results obtained in \cite{Prabhu12,Giorgi11}, where the measurements involved in the calculation of quantum discord are performed over the non-nodal observers. 

The question that naturally arises next is whether a quantum correlation measure, $Q$, which violates one or more of the properties \textbf{P1}-\textbf{P3}, can be monogamous. In \cite{Streltsov12}, it has been shown that a quantum correlation measure, $Q$, which is monogamous, and remains finite when the dimension of one of the subsystems, say, $A$, is fixed, i.e.,
\begin{eqnarray}
 Q(\rho_{AB})\leq f(d_A)<\infty,
 \label{streltsov-eq-1}
\end{eqnarray}
must be zero for all separable states. Here, $d_A$ represents the dimension of $A$, and $f$ is some function.  To prove this, for a generic separable state $\rho_{AB}$, let us consider a symmetric extension of the form  $\rho_{AB_1B_2\ldots B_N}$, where $\rho_{AB}=\rho_{AB_i}\forall i=1,2,\ldots,N$, $N$ being an arbitrary positive integer, such that $Q(\rho_{AB})=Q(\rho_{AB_i})$, $1\leq i\leq N$ \cite{Werner89,Werner89a,Doherty04,Yang06}. This implies $\sum_{i=1}^N Q(\rho_{AB_i})=NQ(\rho_{AB})$. Now, the monogamy of $Q$ implies 
\begin{eqnarray}
 Q(\rho_{A:B_1B_2\ldots B_N})\geq NQ(\rho_{AB}).
 \label{streltsov-eq-2}
\end{eqnarray}
Using  (\ref{streltsov-eq-1}), we have that $Q(\rho_{A|B_1\ldots B_N})$ is finite $\forall N$, including $N\rightarrow\infty$. Therefore, (\ref{streltsov-eq-2}) can be violated with a large enough choice of $N$ if $Q(\rho_{AB})\neq0$ for the separable state $\rho_{AB}$.

\subsection{Relation of entanglement of formation to quantum discord}
\label{quashi-winter}

Let us consider a tripartite pure state $\rho_{ABC}$, which is a purification of the bipartite density matrices $\rho_{AB}$ and $\rho_{AC}$, i.e., $\text{Tr}_{B(C)}[\rho_{ABC}]=\rho_{AC(AB)}$. Let us now consider $I^\leftarrow(\rho_{AC})$, where the only difference with the quantity defined in Eq. (\ref{class-corr}) is that here we assume an optimization over POVMs \cite{Henderson01,Devetak04}. Let $\{p_i,|\psi_i\rangle\}$ be a pure state decomposition of $\rho_{AB}$  that achieves the minimum in the definition of the EoF of $\rho_{AB}$. Now, there must exist a particular measurement setting $\{\tilde{M}_i\}$ corresponding to the subsystem $C$ of the state $\rho_{ABC}$, for which the joint state of the rest of the system, $AB$, turns out to be $|\psi_i\rangle$ with probability $p_i$ corresponding to the $i^{th}$ outcome \cite{Hughston93}. This leaves the subsystem $A$ in the state $\text{Tr}_B(|\psi_i\rangle\langle \psi_i|)$. Following the definition of $I^{\leftarrow}(\rho_{AB})$, this implies \cite{Koashi04} 
\begin{eqnarray}
\label{ap1}
I^{\leftarrow}(\rho_{AC}) &\geq&  S(\rho_A)-\sum_i p_i S\left[\text{Tr}_B(|\psi_i\rangle\langle \psi_i|)\right]\\
\label{ap1a}
&=& S(\rho_A)-E_f(\rho_{AB}).
\end{eqnarray}

One may also consider an alternative approach, where a particular measurement $\{M_i\}$, when performed over the subsystem $C$, attains the maximum in $I^{\leftarrow}(\rho_{AC})=S(\rho_A)-\sum_{i} p_{i} S(\rho_{i})$. In general, $\{M_i\}$ may have a rank more than $1$. It is now suggestive to decompose $\{M_i\}$ into rank-$1$  non-negative operators, $M_{ij}$, satisfying  $M_i= \sum_{j} M_{ij}$. Let us assume that the action of $M_{ij}$ on the subsystem $C$ leaves the subsystem $A$ in $\rho_{ij}$ with probability $p_{ij}$, where the following relations hold: $p_i=\sum_j p_{ij}$ and $\rho_i=\sum_j \rho_{ij}$. Concavity of the von Neumann entropy implies that $S(\rho_A)-\sum_{ij} p_{ij} S(\rho_{ij}) \geq S(\rho_A)-\sum_{i} p_{i} S(\rho_{i})=I^{\leftarrow}(\rho_{AC})$. However, this conflicts with the definition of $I^{\leftarrow}(\rho_{AC})$, unless $S(\rho_A)-\sum_{ij} p_{ij} S(\rho_{ij})=I^{\leftarrow}(\rho_{AC})$. Now consider the action of $\{M_{ij}\}$ on the subsystem $C$. The state, $|\phi_{ij}\rangle$, of the subsystem $AB$, corresponding to the outcome $ij$, is a pure state, and the measurement, $\{M_{ij}\}$, therefore, leads to an ensemble $\{p_{ij},|\phi_{ij}\rangle\}$, satisfying $\sum_{ij} p_{ij}|\phi_{ij}\rangle\bra{\phi_{ij}} = \rho_{AB}$. This also signifies that $\rho_{ij}=\text{Tr}_B\left[|\phi_{ij}\rangle \langle \phi_{ij} | \right]$. Consequently \cite{Koashi04},
\begin{eqnarray}
\label{ap2}
E_f(\rho_{AB}) &\leq&  \sum_{ij} p_{ij} S(\rho_{ij})\\
\label{ap2a}
&=&S(\rho_A)-I^{\leftarrow}(\rho_{AC}).
\end{eqnarray}
Combining (\ref{ap1a}) and (\ref{ap2a}), we have \cite{Koashi04}
\begin{eqnarray}
\label{ap3}
E_f(\rho_{AB})+I^{\leftarrow}(\rho_{AC})=S(\rho_A).
\end{eqnarray}
This directly leads to a relation between entanglement of formation and quantum discord. The above relation turns out to be extremely important to prove several monogamy relations for different quantum correlation measures as can be seen in subsequent sections. However, note that the use of the relation implies that  quantum discord is computed by performing POVMs.

Note here that  $\rho_{ABC}^{\otimes n}$ is a purification of the states, $\rho_{AB}^{\otimes n}$ and $\rho_{AC}^{\otimes n}$. Moreover, the additivity of von Neumann entropy implies that $S(\rho_A^{\otimes n})= n S(\rho_A)$. As a result, one obtains $E_f(\rho_{AB}^{\otimes n})+I^{\leftarrow}(\rho_{AC}^{\otimes n})=n S(\rho_A)$. Dividing both sides by $n$ and taking the limit $n \to \infty$, Eq. (\ref{ap3}) reduces to 
\begin{eqnarray}
\label{ap4}
E_C(\rho_{AB})+C_D^{\leftarrow}(\rho_{AC})=S(\rho_A),
\end{eqnarray} 
where $E_C(\rho_{AB})=\underset{{n \to \infty}}{\text{lim}}\frac{1}{n}E_f(\rho_{AB}^{\otimes n})$ is the entanglement cost for creating the $\rho_{AB}$ by local operations and classical communication (LOCC) from a resource of shared singlets \cite{Hayden01}, and $C_D^{\leftarrow}(\rho_{AC})=\underset{{n \to \infty}}{\lim}\frac{1}{n}I^{\leftarrow}(\rho_{AC})$ is the one-way distillable common randomness of $\rho_{AC}$ \cite{Devetak04}. 

\subsection{Conservation law: Entanglement vs. quantum discord}
\label{subsec:conv-law}

The above discussion directly leads to a conservation relation between EoF and quantum discord of an arbitrary three-qubit pure state, $\rho_{ABC}$. Note here that since $\rho_{ABC}$ is pure, $S(\rho_{A})$ is a good measure of quantum correlation of $\rho_{ABC}$ in the $A:BC$ partition. Eq. (\ref{ap3}) implies that the amount of quantum correlation between the subsystem $A$ and the rest of the system is the sum of the amount of quantum correlation present between $A$ and $B$, and the amount of classical correlation present between $A$ and $C$, thereby imposing a constraint over the distribution of the correlations between $A$ and the rest of the system. Adding the uninterrogated mutual information between $A$ and $C$, given by $\tilde{I}(\rho_{AC})=S(\rho_A)+S(\rho_C)-S(\rho_{AC})$, to both sides of Eq. (\ref{ap3}), one obtains 
\begin{eqnarray}
 E_f(\rho_{AB})=D^\leftarrow(\rho_{AC})+\tilde{S}(\rho_{A|C}).
 \label{koashi-eq-1}
\end{eqnarray}
Proceeding in a similar fashion, one can write $E_f(\rho_{AB})$ and $E_f(\rho_{AC})$ as  
\begin{eqnarray}
 E_f(\rho_{AB})&=&D^\leftarrow(\rho_{BC})+\tilde{S}(\rho_{B|C}),\nonumber \\
 E_f(\rho_{AC})&=&D^\leftarrow(\rho_{AB})+\tilde{S}(\rho_{A|B}).
 \label{koashi-eq-2}
\end{eqnarray}
Since the tripartite state is pure, $E_f(\rho_{C:AB})=S(\rho_C)$ and $E_f(\rho_{B:AC})=S(\rho_{B})$, which, from Eq. (\ref{koashi-eq-2}), implies \cite{Fanchini11}
\begin{eqnarray}
 D^\leftarrow(\rho_{AB})=E_f(\rho_{AC})-E_f(\rho_{C:AB})+E_f(\rho_{B:AC}). \nonumber \\
 \label{fanchini-eq-1}
\end{eqnarray}
Also, noticing that $\tilde{S}(\rho_{A|B})=-\tilde{S}(\rho_{A|C})$ since the state $\rho_{ABC}$ is pure, and using Eqs. (\ref{koashi-eq-1}) and (\ref{koashi-eq-2}), one obtains \cite{Fanchini11}
\begin{eqnarray}
 E_f(\rho_{AB})+E_f(\rho_{AC})=D^\leftarrow(\rho_{AB})+D^\leftarrow(\rho_{AC}).
 \label{fanchini-eq-2}
\end{eqnarray}
Note here that in the above discussion, we have considered the party $A$ to be the nodal observer, and while computing quantum discord, the measurement is performed over the non-nodal observer. The above equation suggests that for an arbitrary tripartite pure state $\rho_{ABC}$, the sum of all possible bipartite entanglements shared by the nodal observer with the rest of the individual subsystems, as measured by the EoFs, can not be increased without increasing the sum of corresponding quantum discords by the same amount. This seems to indicate a ``conservation relation" between EoF and quantum discord with respect to a fixed nodal observer in the case of a given tripartite pure state. 

In this context, we point out that multipartite measures of quantum correlations have been defined as the sum of quantum correlations for all possible bipartitions in a multiparty quantum state, by using EoF, quantum discord, and geometric quantum discord as measures of bipartite quantum correlations. For tripartite pure states, the conservation law implies that certain such multipartite measures corresponding to EoF and quantum discord are equivalent \cite{Ma13,Fanchini12}. In the same vein, Ref. \cite{Daoud13} investigates the above multipartite quantum correlations, in terms of EoFs and quantum discords for even and odd spin coherent states.

Since $D^\leftarrow(\rho_{A:BC})=E_f(\rho_{A:BC})=S(\rho_A)$ for an arbitrary tripartite pure state $\rho_{ABC}$, Eq. (\ref{fanchini-eq-2}) implies an equivalence between the monogamy relation of EoF and quantum discord \cite{Giorgi11}. In the case of mixed states, the conservation law changes into \cite{Fanchini11} 
\begin{eqnarray}
 D^\leftarrow(\rho_{AB})+D^\leftarrow(\rho_{AC})\geq E_f(\rho_{AC})+E_f(\rho_{AB})+\Delta, \nonumber \\
 \label{giorgi-eq-1}
\end{eqnarray}
where $\Delta=S(\rho_{B})-S(\rho_{AB})+S(\rho_{C})-S(\rho_{AC})$. 

\subsection{Relation with interrogated information}
\label{subsec:nec-suff}

We now derive a relation which gives a physical insight into the monogamy property of quantum correlation measures. For an arbitrary tripartite state \(\rho_{ABC}\), an \textit{uninterrogated} conditional mutual information is defined as 
\begin{eqnarray}
\tilde{I}(\rho_{A:B|C}) = \tilde{S}(\rho_{A|C}) - \tilde{S}(\rho_{A|BC}),
\end{eqnarray}
while the corresponding \textit{interrogated} version can be expressed as 
\begin{eqnarray} 
I(\rho_{A:B|C}) = S(\rho_{A|C}) - S(\rho_{A|BC}), 
\end{eqnarray}
involving measurement over one or more of the subsystems. Here, \(\tilde{I}(\rho_{A:B|C})\) and  \(I(\rho_{A:B|C})\) are non-negative, which  is a direct consequence of the non-increasing nature of  conditional entropy with an increase in the number of parties over which it is conditioned. The definitions of $S$ and $\tilde{S}$ are as given in Sec. \ref{subsec:qd}. Given a tripartite quantum state $\rho_{ABC}$, the \textit{interaction information} \cite{Cover06}, \(I(\rho_{ABC})\), is defined as the difference between the information shared by the subsystem \(AB\) when \(C\) is present, 
and when \(C\) is traced out. Since \(\tilde{S}(\rho_{A|C}) = S(\rho_{AC}) - S(\rho_C)\) and \(\tilde{S}(\rho_{A|BC}) = S(\rho_{ABC}) - S(\rho_{BC})\), one can write an  \textit{uninterrogated} interaction information as \cite{Prabhu12}
\begin{eqnarray}
\tilde{I}(\rho_{ABC}) &=& \tilde{I}(\rho_{A:B|C}) - \tilde{I}(\rho_{AB})\nonumber \\ 
&=& S(\rho_{AB}) + S(\rho_{BC}) + S(\rho_{AC}) - S(\rho_{ABC})\nonumber \\
&&-(S(\rho_{A}) + S(\rho_{B}) + S(\rho_{C})). 
\end{eqnarray}

One can also define an \textit{interrogated} interaction information, where the conditional entropies are defined so that a  complete measurement has to be performed  on one of the subsystems. In the case of the tripartite state \(\rho_{ABC}\), an interrogated interaction information is given by
\begin{eqnarray}
I(\rho_{ABC})_{\{\Pi_k^B, \Pi_i^C,\Pi_j^{BC}\}} &=& I(\rho_{A:B|C})_{\{\Pi_i^C,\Pi_j^{BC}\}}  - I(\rho_{AB})_{\{\Pi_k^B\}},\nonumber \\
\end{eqnarray}
where the suffix on \(I(\rho_{ABC})_{\{\Pi_k^B, \Pi_i^C,\Pi_j^{BC}\}}\) indicates that the measurements are performed over \(B\), \(C\), and \(BC\). A similar notation is used to define \(I(\rho_{A:B|C})_{\{\Pi_i^C,\Pi_j^{BC}\}} = S(\rho_{A|C})_{\{\Pi_i^C\}} - S(\rho_{A|BC})_{\{\Pi_j^{BC}\}}\) and \(I(\rho_{AB})_{\{\Pi_k^B\}} = S(\rho_A) - S(\rho_{A|B})_{\{\Pi_k^B\}} \equiv S(\rho_A) - \sum_k p_kS(\rho_{A|k})\). 
For an arbitrary tripartite state \(\rho_{ABC}\), the value of the interrogated interaction information is obtained by performing an optimization over the measurements. One can show that the quantum interaction information \textit{(i)} can be either positive or negative, \textit{(ii)} 
is invariant under local unitaries, and  \textit{(iii)} obeys the inequality  \(I(\rho_{ABC}) \geq \tilde{I}(\rho_{ABC})\) under unilocal measurements, which can be seen directly from the fact that quantum discord is non-negative \cite{Prabhu12}.

If the optimization over the complete set of measurements is performed, the monogamy relation of quantum discord, i.e., 
\begin{eqnarray} 
D^\leftarrow(\rho_{AB})+D^\leftarrow(\rho_{AC})\leq D^\leftarrow(\rho_{A:BC})
\label{monogamy-discord}
\end{eqnarray}
directly leads to  
\begin{equation}
I(\rho_{A:B|C}) - I(\rho_{AB}) \le \tilde{I}(\rho_{A:B|C}) - \tilde{I}(\rho_{AB}).
\label{nec-suff-cond-1}
\end{equation}
On the other hand, assuming the relation (\ref{nec-suff-cond-1}) implies that ${\min}_{\Pi_i^{BC } } S(\rho_{A|BC})_{\{\Pi_i^{BC}\}}$ $-$ $\tilde{S}(\rho_{A|BC})$ $\geq$ $[{\min}_{\Pi_i^B} S(\rho_{A|B})_{\{\Pi_i^B \}}$ $-$ $\tilde{S}(\rho_{A|B})]$ $+$ $[ {\min}_{\Pi_i^C} S(\rho_{A|C})_{\{\Pi_i^C \}}$ $-$ $\tilde{S}(\rho_{A|C})]$, which, in turn, implies the monogamy of quantum discord. Therefore, an arbitrary tripartite quantum state $\rho_{ABC}$ is monogamous with respect to quantum discord if and only if \(I(\rho_{ABC})_{\{\Pi_k^B, \Pi_i^C,\Pi_j^{BC}\}}\leq \tilde{I}(\rho_{ABC})\) \cite{Prabhu12}. For a tripartite pure state, since $\tilde{I}(\rho_{ABC})=0$, the interrogated interaction information is non-positive. 

Consider the space of arbitrary three-qubit pure states, formed by the union of the GHZ and W classes, which are inequivalent under stochastic local operations and classical communication (SLOCC) \cite{Dur00}. The monogamy of quantum discord has been tested numerically for the GHZ and the W classes \cite{Prabhu12}. Evidence for both satisfaction and violation of monogamy relation of quantum discord in the former class has been found, while for the latter, monogamy of quantum discord is always found to be violated \cite{Prabhu12}, if the measurement is performed over non-nodal observers. The violation of monogamy of quantum discord in the case of three-qubit pure states belonging to the W class can be proved analytically using the equivalence of monogamy relations of EoF and quantum discord in the case of tripartite pure states (see Sec. \ref{subsec:conv-law}) \cite{Giorgi11}. Up to local operations, an arbitrary three-qubit pure state can be parametrized as \cite{Acin00,Acin01}
\begin{eqnarray}
\ket{\psi_{ABC}} &=& \lambda_0\ket{000} + \lambda_1e^{i\gamma}\ket{100} + \lambda_2\ket{101}+\lambda_3\ket{110}\nonumber\\&&+\lambda_4\ket{111},
\label{arb-three-qubit-state}
\end{eqnarray}
where $\{\lambda_i:i=1,\ldots,4\}$ and $\gamma$ are real parameters. For $\lambda_4=0$, Eq. (\ref{arb-three-qubit-state}) represents an arbitrary state from the W class, for which the tangle vanishes \cite{Coffman00}, i.e., 
\begin{eqnarray}
 \mathcal{C}^2_{AB}+\mathcal{C}^2_{AC}=\mathcal{C}^2_{A:BC},
 \label{ckw-eq}
\end{eqnarray}
where $\mathcal{C}$ represents the concurrence \cite{Wootters98,Coffman00}. Since $E_f$ $(0\leq E_f\leq 1)$ is a concave function of $\mathcal{C}^2 (0\leq\mathcal{C}^2\leq 1)$, for two-qubit states,  $E_f(\rho_{AB})+E_f(\rho_{AC})\geq E_f(\rho_{A:BC})$. Hence by using Eq. (\ref{fanchini-eq-2}), we obtain the proof of violation of monogamy for quantum discord for the states from the W class. 

\subsection{Monogamy of quantum discord raised to an integer power}
\label{subsec:disc-power}

\begin{figure}
\includegraphics[scale=0.5]{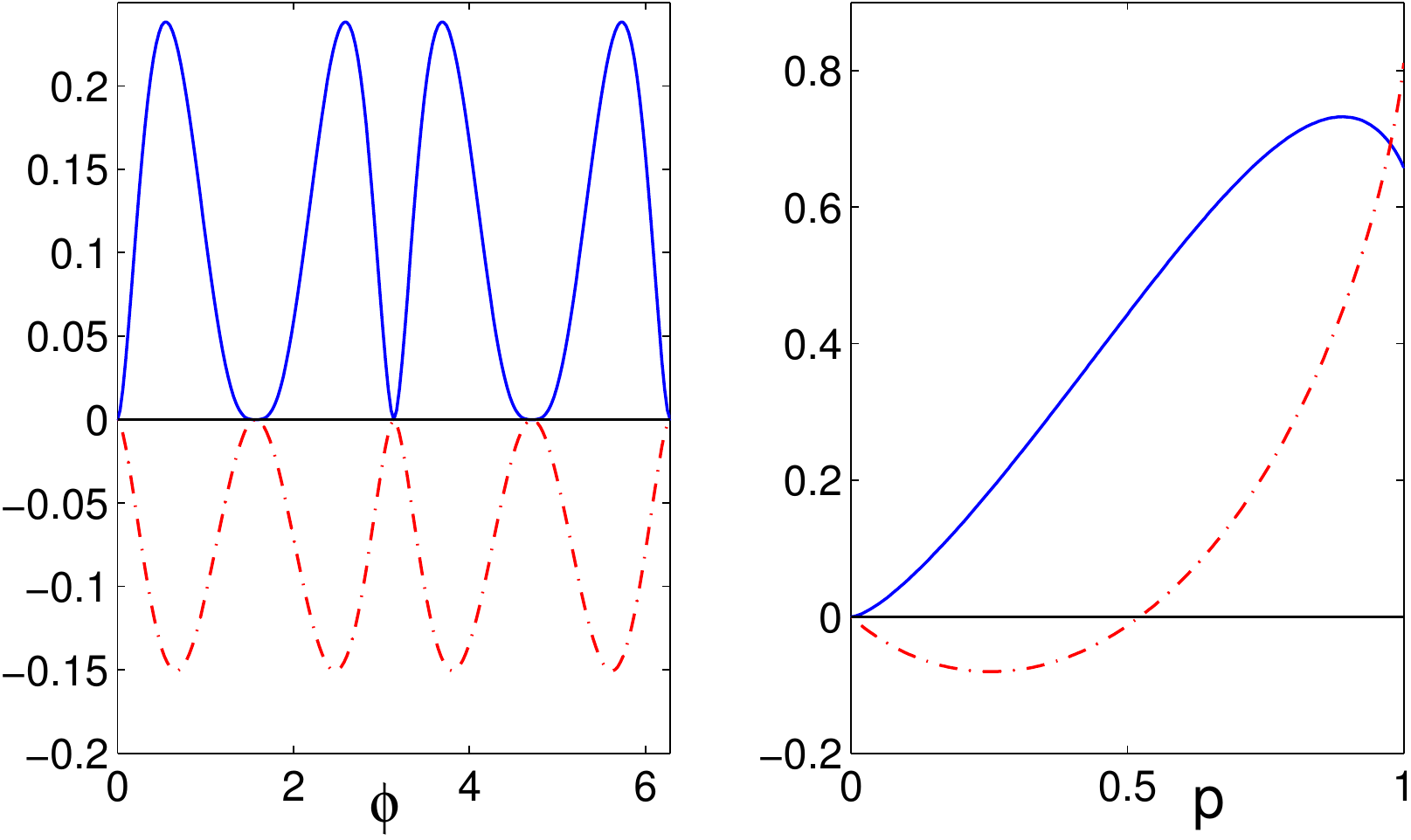}
\caption{(Color online) Variation of the monogamy score for squared quantum discord (blue solid line) in comparison to that of quantum discord (red dash-dotted line) \cite{Bai13}. Left: Variations of the monogamy score for squared quantum discord and quantum discord for the generalized W state, $\ket{\text{gW}}$, as functions of the parameter $\phi$, where the parameter $\theta$ is set to $\pi/4$. Right: Variations of the monogamy score for squared quantum discord and quantum discord in the case of the two-parameter state $\ket{\psi(p,\epsilon)}$, as a function of the state parameter $p$, where the other parameter is chosen to be $\epsilon=0.5$. Reprinted figure with permission from the Authors and the Publisher of Ref.~\cite{Bai13}. Copyright (2012) of the American Physical Society.}
\label{sq-discord}
\end{figure}

Let us consider a bipartite quantum correlation measure, $Q$, which is monotonically decreasing under discarding systems, and remains unchanged under discarding systems only for quantum states satisfying monogamy. Suppose that $\rho_{ABC}$ is a state that violates the monogamy relation for $Q$. This implies 
\begin{eqnarray}
Q(\rho_{A:BC})&<&Q(\rho_{AB})+Q(\rho_{AC}),\nonumber\\
 Q(\rho_{A:BC})&>&Q(\rho_{AB})>0,\nonumber \\
 Q(\rho_{A:BC})&>&Q(\rho_{AC})>0.
\end{eqnarray}
We have additionally assumed that $Q(\rho_{AB})>0$ and $Q(\rho_{AC})>0$. However, the vanishing $Q$ cases can be handled separately. 
These directly lead to 
\begin{eqnarray}
 \lim_{n\rightarrow\infty}\left[\frac{Q(\rho_{AB})}{Q(\rho_{A:BC})}\right]^n&=&0,\nonumber\\
  \lim_{n\rightarrow\infty}\left[\frac{Q(\rho_{AC})}{Q(\rho_{A:BC})}\right]^n&=&0.
\end{eqnarray}
Therefore, for all values of $\epsilon>0$, there exists two positive integers, $n_1(\epsilon)$ and $n_2(\epsilon)$, such that $[Q(\rho_{AB})/Q(\rho_{A:BC})]^m<\epsilon$ for all positive integers $m\geq n_1(\epsilon)$, and similarly for $[Q(\rho_{AB})/Q(\rho_{A:BC})]^m$ with respect to $n_2(\epsilon)$. With a choice of $\epsilon<1/2$, one obtains $[Q(\rho_{AB})/Q(\rho_{A:BC})]^m$, $[Q(\rho_{AC})/Q(\rho_{A:BC})]^m$ $<\epsilon$ for all positive integers $m\geq n(\epsilon)$, where $n(\epsilon)=\max[n_1(\epsilon),n_2(\epsilon)]$, leading to 
\begin{eqnarray}
 \left[\frac{Q(\rho_{AB})}{Q(\rho_{A:BC})}\right]^m+\left[\frac{Q(\rho_{AC})}{Q(\rho_{A:BC})}\right]^m<2\epsilon<1,
\end{eqnarray}
for all positive integers $m\geq n(\epsilon)$. Therefore, considering the bipartite quantum correlation measure $Q^m$, we find that monogamy is obeyed for the state $\rho_{ABC}$ \cite{Salini14}. 

In the case of quantum discord, monogamy property is obeyed for $m\geq 2$ \cite{Bai13,Salini14}, while in the subsequent sections, we shall be discussing similar results regarding other measures of quantum correlations. Note also that if a  quantum correlation measure $Q$ is monogamous in the case of a three-party quantum state, any positive integer power of the measure is also monogamous for the same state, which can be shown directly by expanding $(Q(\rho_{AB})+Q(\rho_{AC}))^m$, and considering the non-negativity of $Q$  \cite{Salini14}. 

We illustrate this by considering the monogamy of squared quantum discord for two specific cases: \textbf{(1)} The three-qubit generalized W state, discussed in Sec. \ref{sec:disc_mono}, and \textbf{(2)} a two-parameter three-qubit pure state, given by $\ket{\psi(p,\epsilon)}=\sqrt{p\epsilon}\ket{000}+\sqrt{p(1-\epsilon)}\ket{111}+\sqrt{(1-p)/2}(\ket{101}+\ket{110})$, where $p$ and $\epsilon$ are real. The variations of the monogamy score of squared quantum discord and quantum discord, as functions of the state parameters, are plotted in Fig. \ref{sq-discord} \cite{Bai13}, which clearly indicates that over the entire range of the state parameters, the squared quantum discord is monogamous, but the quantum discord itself is not.

The case of non-integer powers was considered in \cite{Kumar16b}. It was shown that a monogamous measure remains monogamous on raising its power, i.e., if $Q(\rho_{A:BC})\geq Q(\rho_{A:B})+Q(\rho_{AC})$, then $Q(\rho_{A:BC})^{m}\geq Q(\rho_{A:B})^{m}+Q(\rho_{AC})^{m}$, where $m\geq 1$. Similarly, one can also prove that a non-monogamous measure of quantum correlation remains non-monogamous with a lowering of the power \cite{Kumar16b}. Moreover, it has been pointed out in Ref. \cite{Kumar16b} that if a convex bipartite quantum correlation measure $Q$, when raised to a power $r = 1,2$, is monogamous for pure tripartite states, then $Q^r$ is also monogamous for the mixed states on the given Hilbert space. Note also that all the results mentioned in this subsection, except the monogamy of squared quantum discord, has been generalized to the $N$-partite case \cite{Salini14,Kumar16b}.

\subsection{N-partite quantum states}
\label{subsec:nparty}

Research on the monogamy properties of quantum correlations belonging to either entanglement-separability or the information-theoretic paradigm predominantly deals with tripartite quantum states \cite{Coffman00,Osborne06,Ou07,Prabhu12,Giorgi11}. It is observed that ``good'' entanglement measures \cite{Horodecki00}, which are known to be non-monogamous, in general, for tripartite quantum states, tend to obey monogamy, when considered for quantum states of a moderately large number of parties \cite{Kumar15}. 

Irrespective of the genre of the measure used,  it is possible to determine certain other independent sufficient conditions for an arbitrary bipartite  quantum correlation measure to satisfy monogamy for arbitrary multiparty states. Let us consider an $N$-partite pure state $\rho_{12\ldots N}$, where each of the parties has a dimension $d$. As the first condition, we assume our chosen quantum correlation measure $Q$ to be monogamous for all tripartite quantum states in dimension $d\otimes d\otimes d^m$ with $m\leq N-2$. The monogamy of such an $N$-partite state can be expressed as 
\begin{eqnarray}
 Q(\rho_{1:23\ldots N})\geq Q(\rho_{12})+Q(\rho_{1:34\ldots N}),
 \label{n-party-eq-1}
\end{eqnarray}
where a partitioning of $\rho_{12\ldots N}$, given by $1:2:34\ldots N$, is assumed, and qubit $1$ is chosen to be the nodal observer. 
One may, in turn, partition the state $\rho_{134\ldots N}=\mbox{Tr}_2[\rho_{12\ldots N}]$ as $1:3:4\ldots N$, and can continue to do so 
till the last couple of parties, labeled by $N-1$, and $N$. Recursively  applying the tripartite monogamy relation, (\ref{n-party-eq-1}) can be reduced to 
\begin{eqnarray}
 Q(\rho_{1:23\ldots N})&\geq& Q(\rho_{12})+Q(\rho_{1:34\ldots N}),\nonumber \\
 &\geq& Q(\rho_{12})+Q(\rho_{13})+Q(\rho_{1:45\ldots N}),\nonumber\\
 &\ldots&\nonumber \\
 &\geq&\sum_{k=1}^NQ(\rho_{1k}),
 \label{monogamy_n_party_ineq}
\end{eqnarray}
the intended monogamy of $Q$ for the state $\rho_{12\ldots N}$ \cite{Osborne06,Kumar15}. 

As the second condition, let us assume the convexity of $Q$, and a convex roof definition of $Q$ in the case of a mixed state $\rho_{12\ldots N}$. For a tripartition $1:2:34\ldots N$ of an $N$-party mixed state $\rho_{12\ldots N}$, one obtains 
\begin{eqnarray}
 Q(\rho_{12\ldots N})&=&Q\left(\sum_kp_k(\ket{\psi}\bra{\psi})^k_{12\ldots N}\right),\nonumber \\
 &=&\sum_{k}p_kQ\left(\ket{\psi}^i_{12\ldots N}\right),
 \label{n-party-eq-2}
\end{eqnarray}
where  the optimal convex roof decomposition providing $Q(\rho_{12\ldots N})$  is $\{p_k,\ket{\psi^i_{12\ldots N}}\}$. If one additionally assumes that $Q$ is monogamous for all tripartite pure states in dimension $d\otimes d\otimes d^m$, with $m\leq N-2$, then by using the convexity of $Q$, from Eq. (\ref{n-party-eq-2}), one can write 
\begin{eqnarray}
 Q(\rho_{12\ldots N})\geq Q(\rho_{12})+Q(\rho_{1:34\ldots N}).
\end{eqnarray}
Continuing as before, the monogamy of the mixed state $\rho_{12\ldots N}$ with respect to $Q$, i.e., the relation (\ref{monogamy_n_party_ineq}), can be proven. 

The above result has been numerically tested in Ref. \cite{Kumar15},  by Haar-uniformly generating three-, four-, and five-qubit pure states. For several quantum correlation measures, it is found that the percentages of multiqubit pure states increase with increasing the number of parties. These measures include quantum discord and qantum work deficit. For example, the percentages of three-, four-, and five-qubit states, which are monogamous for quantum discord with measurement performed on the nodal observer,  are respectively 90.5\%, 99.997\%, and 100\%. This result indicates that in the case of a moderately large number of parties in the system,  quantum correlation measures, which are known to be non-monogamous for tripartite quantum states, tend to obey monogamy for almost all states \cite{Kumar15}. The adjective ``almost" is necessary, since for a fixed number of parties, the set of Haar-uniformly generated states may exclude the sets of measure zero in the state space. Therefore, there may exist measure-zero non-monogamous multipartite states which can not be made monogamous for a specified quantum correlation measure by increasing the number of parties. Indeed, it is found that the family of $N$-qubit Dicke states \cite{Dicke54} can not be made monogamous with respect to quantum discord by increasing the number of parties. More specifically, it has been shown that an $N$-partite pure state with vanishing tangle, i.e.,  $\mathcal{C}^2(\rho_{12\ldots N})=\sum_{j=2}^N\mathcal{C}^2(\rho_{1j})$, violates the monogamy relation for quantum discord if the sum of the uninterrogated conditional entropy conditioned on all the non-nodal observers is a negative quantity \cite{Kumar15}. In other words, $\mathcal{C}^2(\rho_{12\ldots N})=\sum_{j=2}^N\mathcal{C}^2(\rho_{1j})$ implies $\delta_{D}^\leftarrow\leq\sum_{j=2}^NS(\rho_{1|j})$,and hence, for pure states having vanishing tangle and $\sum_{j=2}^NS(\rho_{1|j})<0$, the monogamy score for quantum discord is negative, and this is the case for the $N$-qubit Dicke states.

\subsection{Monogamy of quantum discord in open quantum systems}
\label{subsec:mon-open}

Although being of extreme importance, studies of the monogamy property of quantum correlations under noisy environments is limited, possibly due to the mathematical difficulties. Recently, there have been experimental evidence, in photonic systems, of a flow of quantum correlations which occurs between a two-qubit system, $AB$, and its environment, $E$ \cite{Aguilar14}. In this scenario, initially present bipartite entanglement in the system $AB$ decays with time, while multipartite entanglement and multipartite quantum discord emerges in the multipartite system constituted by the bipartite system and its environment. In another work, the dynamics of the monogamy property of quantum discord, in the case of global noise\footnote{The global noisy channel considered corresponds to the completely positive trace preserving (CPTP) map, $\rho \rightarrow \rho^\prime = \zeta (\rho)$, given by $\rho^\prime = \gamma \frac{\mathbb{I}}{d} + (1-\gamma) \rho$, where $\mathbb{I}$ is the identity matrix, $d$ is the dimension of the Hilbert space on which $\rho$ is defined, and $\gamma \in \left[0,1\right]$ is the mixing factor.}, and dissipative and non-dissipative single-qubit quantum channels is discussed in \cite{Kumar16}, by using generalized GHZ \cite{Greenberger89} and generalized W \cite{Zeilinger92,Sen03} states as input states to the noise. As a representative of the dissipative noise, amplitude-damping (AD) channel is used, while phase-damping (PD) and depolarizing (DP) channels are chosen as examples of non-dissipative channels\footnote{The CPTP maps, $\rho\rightarrow\rho^\prime=\zeta(\rho)$, corresponding to these local noisy channels, can be given in the form of their respective Kraus operators, $\{E_k\}$, such that $\rho^\prime=\sum_k E_k\rho E_k^\dagger$, where $\sum_k E_k^\dagger E_k = \mathbb{I}$.   For the single-qubit AD, PD, and DP channels, the Kraus operators are given by $\{E_k^{ad}\}$, $\{E_k^{pd}\}$, and $\{E_k^{dp}\}$, respectively. The Kraus operators for the AD channel are given by 
\begin{eqnarray}
  E_0^{ad}=\left(
  \begin{array}{cc}
     1 & 0\\
     0 & \sqrt{1-\gamma}
  \end{array}
  \right),~~~
  E_1^{ad}=\left(
  \begin{array}{cc}
     0 & \sqrt{\gamma}\\
     0 & 0
  \end{array}
  \right),\nonumber
\end{eqnarray}
while for the PD and the DP channels, they are 
\begin{eqnarray}
E_0^{pd}=\sqrt{1-\gamma}\mathbb{I},~E_1^{pd}=\frac{\sqrt{\gamma}}{2}(\mathbb{I}+\sigma_3), ~E_2^{pd}=\frac{\sqrt{\gamma}}{2},(\mathbb{I}-\sigma_3),\nonumber 
\end{eqnarray}
and
\begin{eqnarray}
E_0^{dp}=\sqrt{1-\gamma}\mathbb{I},~E_i^{dp}=\sqrt{\frac{\gamma}{3}}\sigma_i; i=1,2,3, \nonumber 
\end{eqnarray}
respectively. Here, $\gamma$ is the local noise parameter, with $\gamma\in[0,1]$. }. In case of the three-qubit generalized GHZ state, given by $\ket{\mbox{gGHZ}}=a_0\ket{000}+a_1\ket{111}$, the monogamy score of quantum discord has been shown to be decaying monotonically with increasing strength of the noise parameter. On the other hand, in the case of three-qubit generalized W states given by $\ket{\mbox{gW}}=a_0\ket{001}+a_1\ket{010}+a_2\ket{100}$\footnote{Note that $\ket{\mbox{gW}}$, given in Eq. (\ref{genw}), has been parametrized in the spherical polar co-ordinates, and during numerical simulation, $\theta$ and $\phi$ are generated continuously, while in this case, $a_0$ and $a_1$ are chosen Haar-uniformly to simulate the $\ket{\mbox{gW}}$ state.}, monogamy score for quantum discord exhibits non-monotonic behaviour when the value of the noise parameter is increased. Here, each qubit of the three-qubit states is used as input to the quantum channel under study. For example, when the DP channel is being studied, the three qubits of the three-qubit state are fed into these independent DP channels.  A characteristic value of the noise parameter, called the \emph{dynamics terminal} \cite{Kumar16}, is introduced to quantify the robustness of the monogamy score against a particular type of noise applied to the input state, and depolarized channel is identified as the one that destroys monogamy score faster than the other channels considered. A related statistics of three-qubit states belonging to the sets of generalized GHZ states and generalized W states is obtained numerically, which leads to a conclusive two-step distinguishing protocol to identify the type of noise applied to the three-qubit system. We discuss the details of the two-step channel discrimination protocol in Sec. \ref{subsec:chan-disc}. In \cite{Sarangi14}, dynamics of quantum dissension \cite{Chakrabarty11}, and the monogamy score of quantum discord in the case of amplitude-damping, dephasing, and depolarizing channels are discussed, when the input is from a set of three-qubit states including mixed GHZ states, mixed W states, and a certain mixture of separable and bi-separable states. It was also found there that certain non-monogamous states become monogamous with the increase of noise.

\section{Monogamy of Other Quantum Correlations}
\label{sec:mon-other}

In this section, we discuss the monogamy properties of information-theoretic quantum correlation measures other than quantum discord. We will present results that are specific to the measures, while the results for generic measures of quantum correlations, as discussed in the previous sections, remain valid. We start the discussion with quantum work deficit, which, apart from obeying the properties related to monogamy for general  quantum correlation measures, has a direct relation with the monogamy of quantum discord in the case of tripartite pure states \cite{Salini14}. Assuming that the optimizations for both quantum discord and quantum work deficit of the bipartite state $\rho_{AB}$ takes place for the same ensemble $\{p_k,\rho_{AB}^k\}$, from the definition of quantum discord and quantum work deficit, one can show that
\begin{eqnarray}
 W^\leftarrow(\rho_{AB})=D^\leftarrow(\rho_{AB})-S(\rho_B)+H(\{p_k\}),
\end{eqnarray}
where $H(\{p_k\})$ is the Shannon entropy originating from the local measurement on the party $B$. Since $H(\{p_k\})\geq S(\rho_B)$, 
$W^\leftarrow(\rho_{AB})\geq D^\leftarrow(\rho_{AB})$. This implies that if quantum work deficit is monogamous, i.e., $W^\leftarrow(\rho_{A:BC}) \geq W^\leftarrow(\rho_{AB})+W^\leftarrow(\rho_{AC})$, and since $W^\leftarrow(\rho_{A:BC})=D^\leftarrow(\rho_{A:BC})=S(\rho_{A})$ \cite{Henderson01,Ollivier02,Oppenheim02,Horodecki03,Horodecki05} for pure states, we have 
\begin{eqnarray}
 D^\leftarrow(\rho_{A:BC})=W^\leftarrow(\rho_{A:BC})&\geq& W^\leftarrow(\rho_{AB})+W^\leftarrow(\rho_{AC})\nonumber\\
 &\geq& D^\leftarrow(\rho_{AB})+D^\leftarrow(\rho_{AC}),\nonumber \\
\end{eqnarray}
which implies monogamy of quantum discord. Note that the reverse is not true. Note also that although one can show that $W^\leftarrow(\rho_{AB})+W^\leftarrow(\rho_{AC})\geq D^\leftarrow(\rho_{AB})+D^\leftarrow(\rho_{AC})$ for a three party mixed state under the same assumption of optimization.

In the case of arbitrary three-qubit pure states, quantum work deficit can be both monogamous and non-monogamous. As discussed in Sec. \ref{subsec:disc-power}, similar to quantum discord, a state that is non-monogamous with respect to quantum work deficit can be made monogamous by raising quantum work deficit to an appropriate integer power, $m$. For quantum discord, one requires $m\geq 2$ to obtain monogamy, while for quantum work deficit, the percentage of non-monogamous three-qubit pure states with $m\geq 4$ is approximately $0.22$, implying that a higher integer power of quantum work deficit is necessary to achieve monogamy for almost all three-qubit pure states.

Similar to the findings for quantum discord and quantum work deficit, and from the discussion in Sec. \ref{sec:disc_mono}, geometric quantum discord is not monogamous in general, since it is non-zero in the case of some separable states. However, in contrast to quantum discord and quantum work deficit,  geometric quantum discord is always monogamous for an arbitrary three-qubit pure state $\rho_{ABC}$ \cite{Streltsov12}. Note that the geometric quantum discord, in the present case, is computed by performing the minimization over the set of all ``classical-quantum states" instead of ``quantum-classical states'' as defined in Sec. \ref{subsec:gqd}. This is proved by showing the existence of a classical-quantum state, $\sigma_{ABC}$, for which $D_{G}(\rho_{A:BC})\geq||\rho_{AB}-\sigma_{AB}||^2_2+||\rho_{AC}-\sigma_{AC}||^2_2$, where $\sigma_{AB(AC)}=\mbox{Tr}_{C(B)}(\sigma_{ABC})$, whenever $\rho_{ABC}$ is pure. Since the right hand side of the inequality is always bigger than $D_G(\rho_{AB})+D_G(\rho_{AC})$, due to the minimization involved in geometric quantum discord, we obtain the claimed monogamy of $D_G$. Monogamy of geometric quantum discord is also considered in Ref. \cite{Song13a}, while in the multiqubit scenario, the monogamy of $D_G$ is addressed in Refs. \cite{Bai14b,Rana12}. Ref. \cite{Daoud15} investigates the monogamy of geometric quantum discord in photon added coherent states. 

The monogamy of quantum correlations in the case of three-qubit pure symmetric states, in the Majorana representation \cite{Majorana32}, was addressed in \cite{Sudha12}, where  the Rajagopal-Rendell quantum deficit (RRQD) \cite{Rajagopal02,Devi08} is used as the quantum correlation measure. For a bipartite quantum state $\rho_{AB}$, RRQD is defined as the relative entropy distance \cite{Wehrl78,Vedral02} of the state $\rho_{AB}$ from $\rho^d_{AB}=\sum_{ij}p_{ij}\ket{i}\bra{i}\otimes \ket{j}\bra{j}$, which is diagonal in the eigenbasis of the marginals, 
$\rho_A$, and $\rho_B$, given by $\{\ket{i}\}$ and $\{\ket{j}\}$, respectively. Here, $p_{ij}=\langle j|\langle i|\rho_{AB}|i\rangle|j\rangle$, with $\sum_{i,j}p_{ij}=1$. Mathematically, 
\begin{eqnarray}
 D_{RR}=S(\rho_{AB}||\rho_{AB}^d),
\end{eqnarray}
where for two arbitrary density matrices $\rho$ and $\sigma$, $S(\rho||\sigma)=\mbox{Tr}(\rho\log_2\rho-\rho\log_2\sigma)$. In the case of three-qubit pure symmetric states, RRQD was shown to be non-monogamous in general \cite{Sudha12}.  In particular, it was shown that although generalized W states can satisfy as well as violate the monogamy inequality for RRQD, the generalized GHZ states always satisfy the relation.

We conclude this section by mentioning the monogamy property of measurement induced non-locality, introduced by Luo and Fu \cite{Luo11}, and defined as 
\begin{eqnarray}
 N(\rho_{AB})=\underset{\{\Pi_A\}}{\max}||\rho_{AB}-\rho_{AB}^\prime||^2_2,
\end{eqnarray}
where $\rho^\prime_{AB}=\sum_i\Pi_A^i\otimes\mathbb{I}_B\rho_{AB}\Pi_A^i\otimes\mathbb{I}_B$, $\mathbb{I}_B$ is the identity matrix in the HIlbert space of $B$, and $\{\Pi_A^i\}$ is the set of elements of a projective measurement for which $\sum_i\Pi_A^i\rho_{A}\Pi_A^i=\rho_A$. For this measure, three-qubit pure states belonging to the GHZ and the W class can be non-monogamous in general \cite{Sen12}, although unlike quantum discord, both the generalized GHZ and generalized W states satisfy the monogamy relation.

\section{Relation with other multiparty measures}
\label{sec:oth-meas}

An important perspective of the monogamy inequality of quantum correlations can be obtained by harvesting its inherent multipartite nature, and establish relations between the monogamy scores and the other quantum correlation measures including bipartite and multipartite entanglement.  In several works, monogamy scores of a given quantum correlation measure has been used as important markers of multipartite quantum correlations. In the seminal paper  by Coffman, Kundu, and Wootters \cite{Coffman00} introducing the monogamy inequality for tripartite states, tangle has been described as ``essential three-qubit entanglement''. Over the years, quantities such as monogamy scores for different quantum correlation measures, including quantum discord and quantum work deficit,  have been used as bona-fide measures of multipartite quantum correlations \cite{Bera12,Prabhu12}.

Recently, the relations of monogamy score for quantum discord with different multipartite quantum correlation measures, such as tangle \cite{Coffman00}, genuine multiparty entanglement measures quantified by generalized geometric measure (GGM) \cite{Sen10, Biswas14}, and global quantum discord \cite{Rulli11},  have been established \cite{Prabhu12,Braga12,Liu14}. Significant approaches to quantify multipartite entanglement and quantum correlations using the monogamy principle have also been undertaken \cite{Bai13,Bai14,Bai14a,Liu16}. Monogamy scores of quantum discord and violation of Bell inequalities have also been connected \cite{Pandya15, Sharma16}. Tripartite dense coding capacities are also shown to have relations with the monogamy score of quantum discord \cite{Nepal13, Das14}.

\subsection{Monogamy score versus other multi-site quantumness measures}
\label{subsec:mono-ggm}

In Ref.~\cite{Kumar15a}, the monogamy score of a quantum correlation measure for an $N$-qubit pure multipartite state, $\rho_{12\ldots N}=\ket{\Psi}\bra{\Psi}$, is shown to be intrinsically related to the genuine multipartite entanglement, as quantified by the GGM \cite{Sen10,Biswas14}. An $N$-party pure quantum state $\ket{\Psi}$ is genuinely multipartite entangled if there exist no bipartition across which the state is product. The GGM, $\cal{G}$, of $|\Psi\rangle$ is defined as
\begin{equation}
\cal{G}(|\Psi\rangle) = 1 - \max_{\left\{|\Phi\rangle\right\}}|\langle \Phi|\Psi \rangle|^2, 
\end{equation}
where the maximization is over the set of states $\left\{|\Phi\rangle\right\}$, which  are not genuinely multiparty entangled. The GGM is known to be an entanglement monotone \cite{Sen10}. The above expression for GGM can be simplified to the form, $\cal{G} = 1 - \max_{ k \in [1,N/2]}  
\left[\left\{\xi_{m}(\rho^{(k)})\right\}\right]$, where $\left\{\xi_{m}(\rho^{(k)})\right\}$ is the set of highest eigenvalues of all possible reduced $k$-qubit states, where $k$ ranges from $1$ to $N/2$.

For three-qubit pure states, the maximum eigenvalue has to arise from single-qubit density matrices. Hence, an immediate relation between the 
monogamy score $\delta_{Q}$ and the GGM can be established in this case \cite{Prabhu12,Prabhu12a}. Let $a =\max \{\xi_{m}(\rho^{(1)})\}$ be the maximum eigenvalue corresponding to the single-qubit reduced state, $\rho^{(1)}$ of $|\Psi\rangle$, so that $\cal{G}$ = $1 - a$. From Eq. (\ref{score_nparty}), $\delta^{j}_{Q} \leq {Q}(\rho_{j:\mathrm{rest}})$, where $j$ is the nodal qubit. Now,  the quantity ${Q}(\rho_{j:\mathrm{rest}})$ is a function of $a$, say ${F}^{{Q}}(a)$. Moreover, we have $\cal{G}$ = $1 - a$, which gives us ${F}^{{Q}}(a)$ = ${F}^{{Q}}(1-\cal{G})$, which gives us the bound, $\delta^{j}_{Q} \leq {F}^{{Q}}(a) = {F}^{{Q}}(1-\cal{G})$. Now the monogamy score, $\delta_{Q}$, is defined as the minimum score over all possible nodes, implying $\delta_{Q} \leq \delta^{j}_{Q}$. Hence, we obtain an upper-bound on the monogamy score in terms of a function of GGM, given by
\begin{equation}
\delta_{Q}(|\Psi\rangle) \leq  {F}^{{Q}}(1-\cal{G}(|\Psi\rangle)).
\end{equation}
For quantum correlation measures that reduce to the von Neumann entropy for pure states, such as distillable entanglement \cite{Bennett96}, entanglement cost \cite{Bennett96b}, entanglement of formation \cite{Bennett96a}, squashed entanglement \cite{Koashi04, Christandl04}, relative entropy of entanglement \cite{Vedral97, Vedral02}, quantum discord \cite{Henderson01, Ollivier02}, and quantum work-deficit \cite{Oppenheim02, Horodecki03, Horodecki05, Devetak05}, the function   ${F}^{{Q}}(1-\cal{G})$ = $h(\cal{G})$, where $h(x)$ is the Shannon entropy. For the entanglement monotones, squared concurrence \cite{Coffman00} and squared negativity \cite{Vidal02, Ou07}, the quantity ${F}^{{Q}}(1-\cal{G})$ is equal to  $z\cal{G}(1-\cal{G})$, where $z = 4$ and $1$ for $\cal{C}^2$ and $\cal{N}^2$, respectively. 
We present a plot in Fig. \ref{light-cone} of the upper bound for the case of three-qubit pure states, with the quantum correlation being chosen as quantum discord.
The above relation can be generalized to $N$-qubit pure states, where the upper bound on $\delta_{Q}$ in terms of the entropic or quadratic functions of GGM, can be shown to exist for all states that satisfy a set of necessary conditions. For instance, for all $N$-qubit states $|\Psi\rangle$ with $\cal{G}$ = $1 - a$, the upper bound is universally valid. In other words, for $N$-qubit state $|\Psi\rangle$, if the maximum eigenvalue, among eigenvalues of all local density matrices, is obtained from a single-qubit density matrix, it has been shown that the upper-bound remains valid \cite{Kumar15a}.

An important implication of the above bound is that it is even for those quantum correlation measures for which the corresponding $\delta_{Q}$ can not be explicitly computed for arbitrary states. Examples of such measures include distillable entanglement, entanglement cost, and relative entropy of entanglement. The theorem implies that any possible value  for these measures will always result in a $\delta_{Q}$ that lies on or above the boundary.

Apart from entanglement measures, violations of Bell inequalities are important indicators of quantumness present in compound systems \cite{Clauser69}.  The two-point correlation function Bell inequality violations, and their monogamy properties \cite{Kurzynski11,Toner09} have been connected with the monogamy scores of entanglement and quantum discord in three-qubit quantum systems \cite{Pandya15, Sharma16}. It has been shown that for three-qubit pure states, the monogamy scores for quantum correlations including quantum discord can be upper-bounded by a function of the monogamy score for Bell inequality violation \cite{Pandya15}. Moreover, it was shown in Ref. \cite{Sharma16} that if the monogamy score for quantum discord in the case of an arbitrary three-qubit pure state, $|\Psi\rangle$,  is the same as that of the three-qubit generalized GHZ state, then the monogamy score corresponding to the Bell inequality violation of $\ket{\Psi}$  is bounded below by the same as that of the generalized GHZ state, when the measurements in quantum discord are performed on the non-nodal observer. In case the measurements are performed on the nodal observer, the role of the generalized GHZ state is replaced by the ``special" GHZ state of $N$ qubits \cite{Sharma16}, given by $\ket{\mbox{sGHZ}}=\ket{00\ldots 0}_N+\ket{11}\otimes(\beta\ket{00\ldots 0}+\sqrt{1-\beta^2}e^{-i\theta}\ket{11\ldots 1})_{N-2}$, where $\beta\in[0,1]$, and $\theta$ is a phase.

\subsection{Information complementarity: Lower bound on monogamy violation}
\label{subsec:mono-info-compl}

For a multipartite system in a pure quantum state, $\rho_{12\ldots N}$, let us first divide the whole system into two parts, $x$ and $y$, such that $x\cup y=\{1,2,\ldots,N\}$. It is possible to derive an information-theoretic complementarity relation between the purity of the subsystem $\rho_x$, where $\rho_x$ = $\textrm{Tr}_y\rho_{12\ldots N+1}$, and the bipartite quantum correlation shared between the subsystem $x$ with the rest of the system, i.e., with $y$, and is given by  \cite{Bera16}
\begin{equation}
\label{eq:cr}
{
 P(\rho_{x}) + \mathcal{Q}(\rho_{xy}) \leq b %\left.\mathrel{}\middle|\mathrel{}\right.
        \begin{cases}
        = 1,  & \text{if } d_{x} \leq d_{y}, \\
        = 2 - \frac{\log_2 d_y}{\log_2 d_{x}}, & \text{if } d_{x} > d_{y},
        \end{cases}}.
\end{equation}  
Here, $d_{x(y)}$ is the Hilbert-space dimension of $x(y)$, $P$ = $\frac{\log_2 d_{x}-S(\rho_{x})}{\log_2 d_{x}}$ is the normalized purity of the subsystem $x$, $S(\rho_{x})$ is the von Neumann entropy of $\rho_x$, and $\mathcal{Q}(\rho_{xy})=\frac{{Q}(\rho_{xy})}{\min \{\log_2 d_{x}, \log_2 d_{y}\}}$ is the normalized quantum correlation (with $Q(\rho_{xy})$ being the corresponding quantum correlation) shared between the subsystems $x$ and $y$. The proof of the relation (\ref{eq:cr}) requires that $Q(\rho_{xy})$ satisfies the conditions $Q(\rho_{xy}) \leq S(\rho_{x})$. However, it is independently satisfied by several important quantum correlation measures \cite{Bera16, Kumar16a}. 
The above complementarity relation has useful application in quantum key distribution \cite{Bera16}.

If we now consider a non-monogamous normalized bipartite quantum corelation measure, $\mathcal{Q}$, which could, for example, be normalized quantum discord or normalized quantum work deficit,  we can obtain a useful lower bound on the monogamy score in terms of purity. For an $N$-qudit state, using (\ref{score_nparty}) and (\ref{eq:cr}), one obtains the relation \cite{Kumar16a},
\begin{equation}
\delta_{\mathcal{Q}} \geq -(N-2)\left(1 - {P}({\rho_{n_0}}) + \frac{1- x_0}{N-2} \right),
\end{equation}
where $n_0$ is the nodal qubit and $x_0$ = ${P}({\rho_{n_0}})$ + $\mathcal{Q}({\rho_{n_0:\textrm{rest}}})$. For $x_0 \geq 1$ and large $N$, we obtain $\delta_{\mathcal{Q}} \geq -(N-2)(1 - {P}({\rho_{n_0}}))$, which provides a nontrivial lower bound for the monogamy score of ${Q}$. For three-qubit states, the lower bound of $\delta_{\mathcal{Q}}$ reduces to $\delta_{\mathcal{Q}} \geq -S(\rho_{n_0})$.
See Fig. \ref{light-cone}.

\subsection{Relation with multiport dense coding capacity}
\label{subsec:mono-dense}

Another application of the monogamy inequality and the monogamy score is their role in estimating optimal classical information transfer in multiport dense coding protocols \cite{Bennett92,Bose00,Hiroshima01,Ziman03,Bruss04,Bruss06}. A complementarity relation of the monogamy score for quantum correlation measures, such as squared concurrence and quantum discord, with the maximal dense coding capacity for pure tripartite quantum states was established in Ref.~\cite{Nepal13}. The dense coding capacity of a bipartite quantum state $\rho_{12}$ is given by \cite{Bose00,Hiroshima01,Ziman03,Bruss04,Bruss06}
\begin{equation}
\mathbb{C}(\rho_{12}) = \max\left[\log_2 d_1,\log_2 d_1 + S(\rho_{2})- S(\rho_{12})\right].
\end{equation}
Without a shared entangled resource, the capacity would be $\log_2 d_1$, and hence the quantum advantage is $\mathbb{C}_{\mathrm{adv}} =\max \{0,S(\rho_{2})- S(\rho_{12})\}$. One can consider a multiport communication with $\rho_{12\ldots N}$ as the resource state, where $1$ is the sender, and the rests are the receivers. The quantum advantage in  multiport dense coding for the transfer of classical information from $1$ to $N-1$ individuals  can be defined as $\mathbb{C}_{\mathrm{adv}} = \max\left[\{S(\rho_{i})-S(\rho_{1i})|\forall~ i=2,\ldots,N\}, 0\right]$. Now, if one considers the set of pure tripartite quantum states, the monogamy score for squared concurrence and quantum discord are intrinsically related to the quantum advantage in dense coding. Specifically, it was shown that for any  fixed monogamy score, the maximum quantum advantage is obtained from a single parameter family of three-qubit pure states, given by $\ket{\psi_\alpha}=\ket{111}+\ket{000}+\alpha(\ket{101}+\ket{010})$ \cite{Nepal13}. 
%For $\ket{\psi_\alpha}$, $\mathbb{C}_{\mathrm{adv}}$ is given by \cite{Nepal13}  
%\begin{eqnarray}
%\mathbb{C}_{\mathrm{adv}} (|\psi_{\alpha}\rangle) &=& 
%1 + \left(\frac{1-2\alpha}{2(1 +\alpha^2)}\right) \log_2 \left(\frac{1-2\alpha}{2(1 +\alpha^2)}\right)\nonumber \\
%& &+\left(\frac{1+2\alpha}{2(1 +\alpha^2)}\right) \log_2  \left(\frac{1+2\alpha}{2(1 +\alpha^2)}\right),
%\label{eq:dca}
%\end{eqnarray}
It is possible to derive a complementarity relation between the monogamy score of quantum discord and the quantum advantage. A similar complementarity relation exists between the monogamy score of squared concurrence and $\mathbb{C}_{\mathrm{adv}}$.

Monogamy scores of entanglement and quantum discord have also been related to the multiparty dense coding capacity between several senders and a single receiver \cite{Das14}. In particular, it was shown that in the noiseless scenario, among all multiqubit pure states with an arbitrary but fixed multiparty dense coding capacity, the generalized GHZ state has the maximum monogamy score for quantum discord, i.e., if $\tilde{C}(|\psi\rangle) = \tilde{C}(|gGHZ\rangle)$, it implies $\delta_D(|\psi\rangle) \leq \delta_{D}(|gGHZ\rangle)$, where $\tilde{C}(\rho_{12\ldots N})=\log_2 d_{1 \ldots N-1}+S(\rho_N)-S(\rho_{1 \ldots N})$, with $\rho_{1 \ldots N}$ being a state shared between $N-1$ senders, $1,2, \ldots, N-1$, and the receiver, $N$. Here, $d_{1\ldots N-1}=d_1d_2\ldots d_{N-1}$. We have suppressed the arrow in the superscript of $\delta_{D}$ here, as the result is true independent of the direction of the arrow. The above result is also true if $\delta_D$ is replaced by the tangle. Note also that Ref. \cite{Das14} also considers the noisy channel case.

\section{Physical applications}
\label{sec:mono-appli}

In recent years, several works have been undertaken to elucidate the role of monogamy of quantum correlations in studying quantum 
systems and their dynamics. In particular, the concept of monogamy has been used to characterize quantum states \cite{Giorgi11,Prabhu12, Sudha12, Sen12} and channels \cite{Kumar16}, and also to provide deeper understanding of many physical properties such as critical phenomena in many-body systems \cite{Allegra11,Song13,Qiu14,Qin16,Rao13} involving complex quantum models such as frustrated spin lattices \cite{Rao13}, and  biological compounds \cite{Zhu12,Chanda14}. Moreover, monogamy also provides an important conceptual basis to quantify quantum correlations in multiparty mixed states,  by using the concept of the monogamy score in situations where the usual measures of quantum correlations are neither easily accessible nor computable. More precisely, given a bipartite quantum correlation measure, the monogamy score for that measure defined for a given multiparty system, leads us to a measure of multipartite quantum correlation, without an increase in the complexity on both experimental and theoretical fronts as compared to those at the level of the bipartite measure.

\subsection{State discrimination}

\begin{figure}
 \includegraphics[scale=0.6]{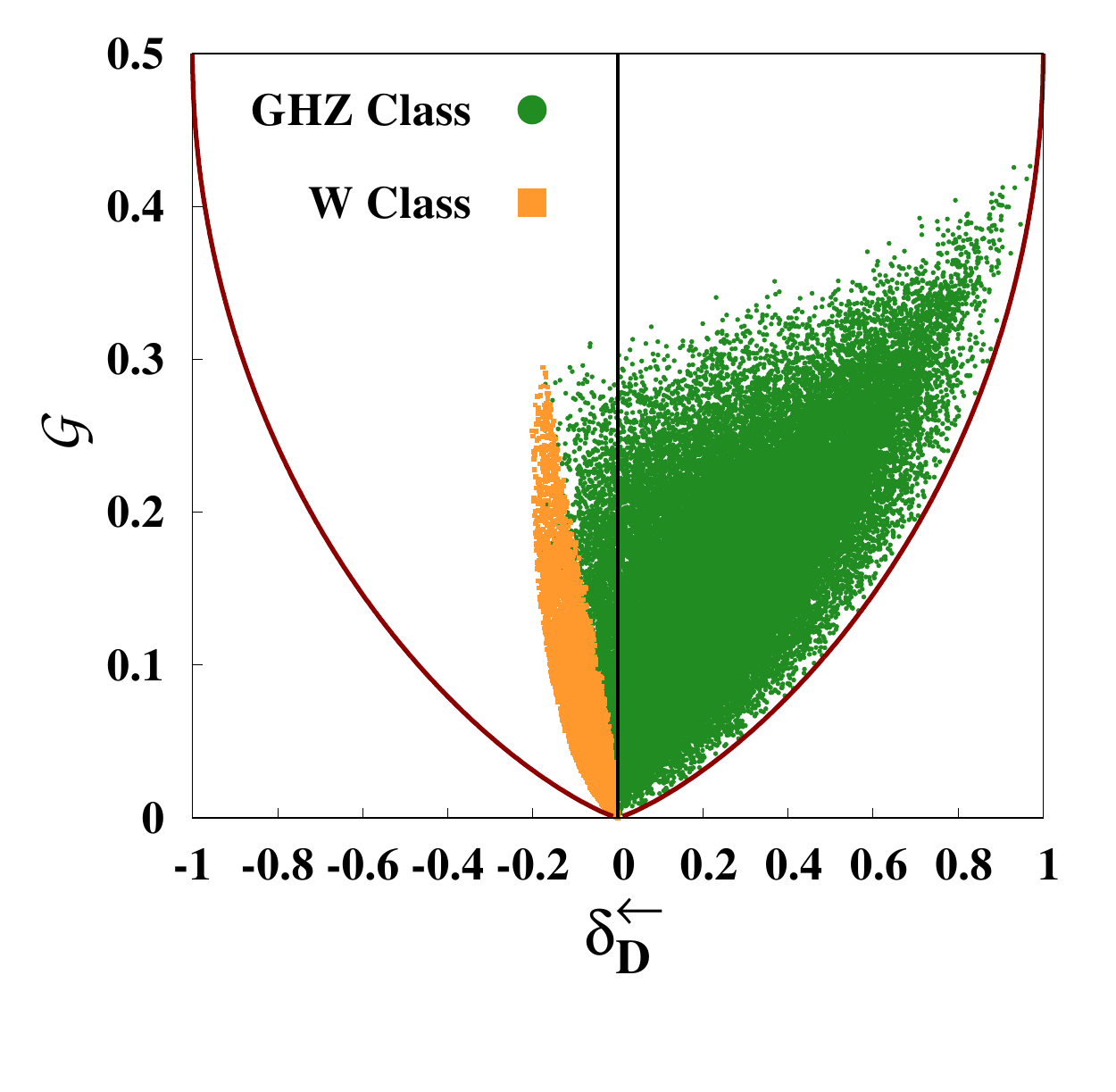}
 \caption{(Color online) Upper and lower bounds on the monogamy score of quantum discord for pure three-qubit GHZ- and W-class states. For three-qubit pure states, the upper bound using the GGM, $\mathcal{G}$, and the lower bound from the information complementarity, are given by the quantity $h(\mathcal{G})$ and $-h(\mathcal{G})$, respectively, where $h$ is the binary entropy function. 
The scatter points in the figure correspond to \(10^6\) three-qubit pure states generated Haar uniformly. An equal number of W-class states are also Haar uniformly generated.  
 The figure shows that all W-class states have a negative monogamy score that is weakly bounded below by $-S(\mathcal{G})$, whereas GHZ-class states can have both positive and negative monogamy scores.  Reprinted figure with permission from the Authors and the Publisher of Ref. \cite{Prabhu12a}. Copyright (2012) of the American Physical Society.}
 \label{light-cone}
\end{figure}

An important aspect in the study of the monogamy properties of quantum correlation measures, such as quantum discord and quantum work-deficit, is that these measures are not universally monogamous. In other words, for these quantum correlation measures, the monogamy inequality is not universally satisfied for all quantum states. This dichotomy allows the monogamy of quantum correlations to be an important figure of merit in state discrimination. In particular, for three-qubit pure states, it was shown that while the generalized GHZ states are always monogamous with respect to quantum discord, all generalized W states violate the monogamy inequality \cite{Prabhu12, Giorgi11}.  The above results were extended to the more general sets, viz. the GHZ and the W class states, where it was shown that more than $80\%$ of the Haar-uniformly generated GHZ class states satisfy monogamy, in contrast to $W$ class states which are always non-monogamous (see Fig. \ref{light-cone}). The monogamy score, therefore, plays a role that is akin to entanglement witnesses \cite{Horodecki96,Terhal01,Guhne02,Lewenstein02,Bruss02,Guhne03}. Indeed a given linear entanglement witness allots values (real numbers) with a certain sign (say, negative), to all separable states while for entangled states, the same witness can have values of both signs. So, a positive value of the witness for a certain state immediately implies that the state is entangled. Similarly, a positive value of the monogamy score for quantum discord for a three-qubit pure state implies that the state is from the GHZ class.  Subsequently, the comparative studies of the SLOCC inequivalent classes was discussed using the monogamy score of another measure of quantum correlation \cite{Sudha12}, namely, the quantum deficit \cite{Rajagopal02, Devi08}. It was shown that while generalized W states may violate monogamy, generalized GHZ states always satisfy the monogamy inequality of quantum deficit. We therefore find that the state discrimination protocol using monogamy inequalities is dependent on the choice of the quantum correlation. For instance, using the monogamy properties of measurement induced non-locality \cite{Luo11}, it was shown that both tripartite generalized GHZ and generalized W states are monogamous\cite{Sen12}.

The monogamy inequality and the related monogamy scores have also been used to characterize pure tripartite quantum states \cite{Bera12},  by finding the relation of the monogamy scores for those states with the corresponding values for measures of genuine multipartite entanglement, viz. the GGM \cite{Sen10, Biswas14} and the multipartite Mermin-Klyshko Bell inequalities \cite{Bell64, Mermin90, Ardehali92,Belinskii93}. In particular, tripartite states that have a vanishing monogamy score for quantum discord have been explored in this way \cite{Bera12}. Some of these aspects have been discussed in Sec. \ref{sec:oth-meas}.

\subsection{Channel discrimination}
\label{subsec:chan-disc}

\begin{figure}
 \includegraphics[scale=0.3]{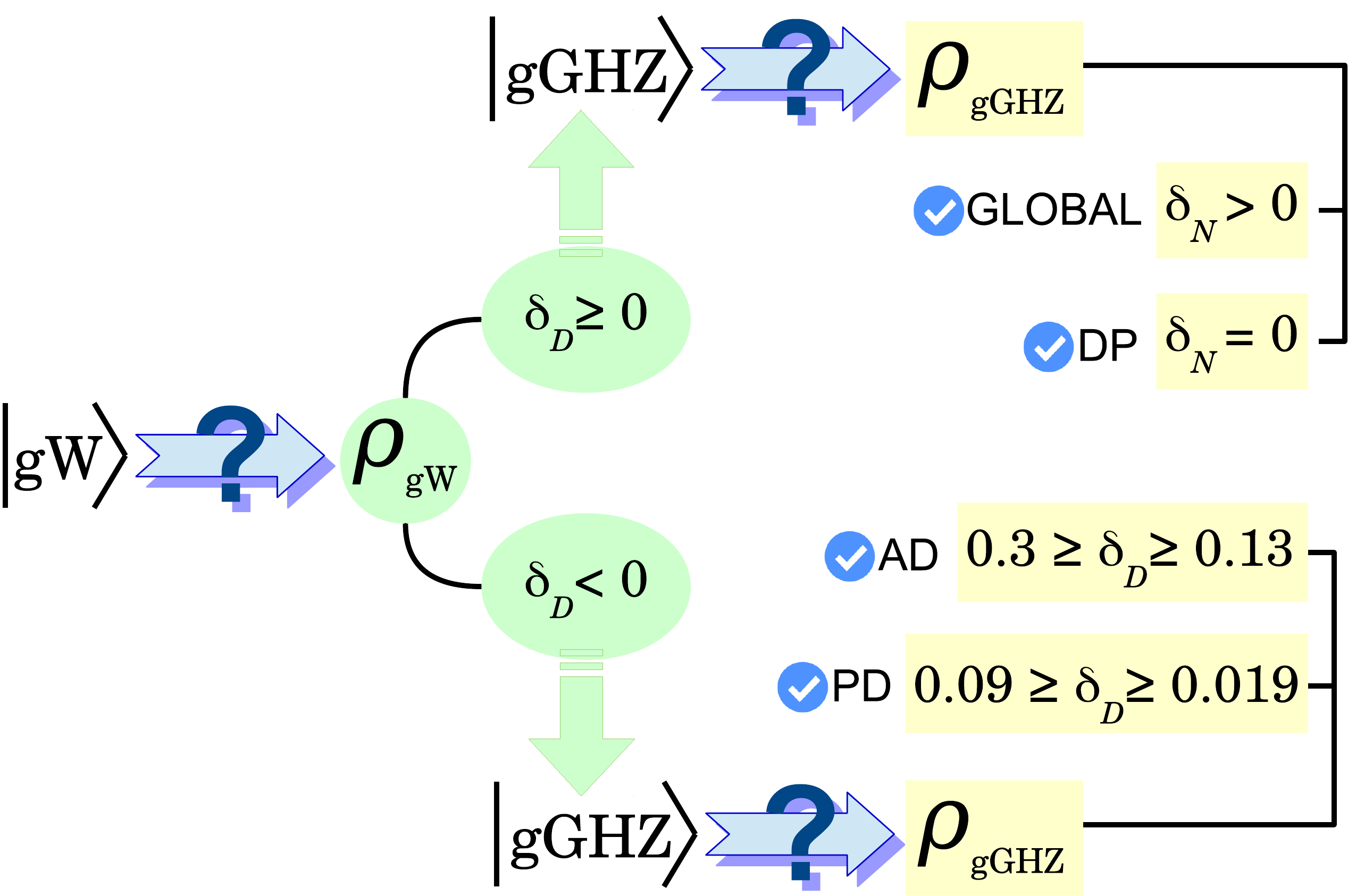}
 \caption{(Color online) Schematic representation of the two-step channel discrimination protocol proposed in \cite{Kumar16}. Reproduced figure with permission from the Authors and the Publisher of Ref. \cite{Kumar16}. Copyright (2016) of Physics Letter A (Elsevier).}
 \label{chan-discr}
\end{figure}

Another application of the monogamy considerations of quantum correlations comes from the study of their behavior under the action of global and local noisy channels. It has been observed that an analysis of the dynamics of  the monogamy scores of quantum discord and entanglement, quantified by negativity, for initial tripartite generalized GHZ and generalized W states can conclusively identify the noisy channel acting on the system \cite{Kumar16}. By analyzing the monogamy scores for quantum discord ($\delta_{D}$) and negativity ($\delta_{\mathcal{N}}$), with generalized GHZ and generalized W states as inputs, a two-step discrimination protocol to identify the above channels has been developed. To describe the protocol, let us consider an apriori unknown  noisy channel, chosen from a set containing a global noise channel, and the AD, PD, and DP channels. See Sec. \ref{subsec:mon-open} for descriptions of these channels. In step $1$ of the channel-discrimination protocol, one feeds generalized W state to the unknown channel with moderate noise, i.e., $\gamma \in \left[0.4,0.6\right]$. The monogamy score of discord is the primary indicator in this step. For the global noise and DP channel, $\delta_{D} \geq 0$, while it is strictly negative for the AD and PD channels. In step $2$ of the protocol, the same unknown channel,  with moderate noise, is applied to an generalized GHZ state, and in this instance, both $\delta_{\mathcal{N}}$ and $\delta_{D}$ of the output state are estimated. It is observed that $\delta_{\mathcal{N}} >$ 0 for the global noisy channel, but vanishes for the DP channel. Note that step $2$ distinguishes between the instances which exhibit $\delta_{D} \geq$ 0 in step 1. On the other hand, in step 2, $\delta_{D} \geq 0.13$ for the AD channel and $\delta_{D} \leq 0.09$ for the PD channel. Therefore, the values of $\delta_{D}$ and $\delta_{\mathcal{N}}$ together can discriminate between the global noise and the three local noisy channels \cite{Kumar16}. A schematic representation of the two-step channel discrimination protocol can be found in Fig. \ref{chan-discr}.

Identification of quantum channels using the dynamics of monogamy scores of quantum correlations is potentially an important addition to the literature on channel identification \cite{Fujiwara01} and estimation \cite{Sarovar06}, which is a significant yet less discussed part of the vast literature on quantum state estimation \cite{Paris04}. 

\subsection{Characterization of quantum many-body systems}

\begin{figure}
 \includegraphics[scale=0.9]{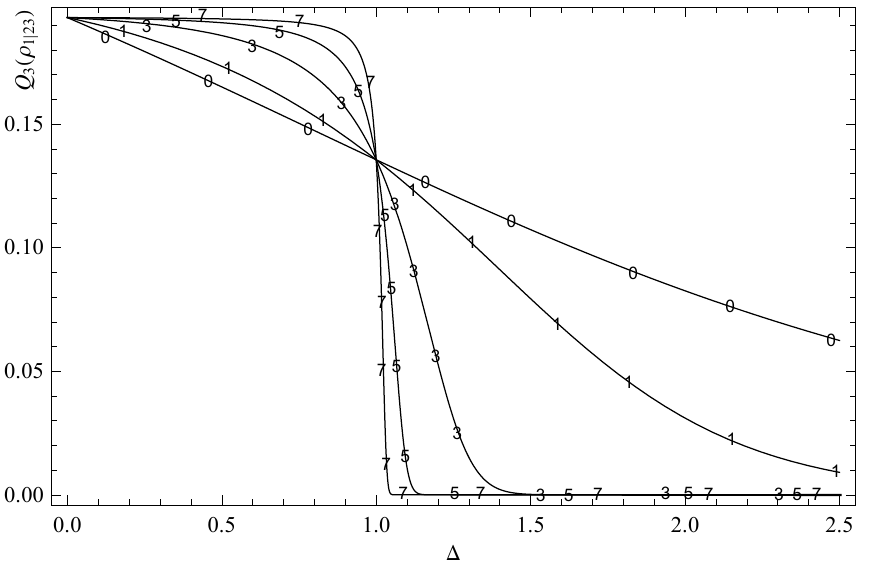}
 \caption{(Color online) Variation of the monogamy score of squared quantum discord as a function of $\Delta$ with increasing 
 system-size in the case of the XXZ model. Reprinted figure with permission from the Authors and the Publisher of Ref. \cite{Qiu14}. Copyright (2014) of the Europhysics Letters (Institute of Physics).}
 \label{mono_qpt}
\end{figure}

Although several studies have attempted to characterize many-body quantum systems using quantum correlations beyond entanglement (for a review, see Ref.~\cite{Modi12}), those engaging monogamy of quantum correlations to understand cooperative properties in strongly-correlated many-body systems are relatively scarce. In Ref.~\cite{Allegra11}, monogamy property of quantum discord is used to characterize the ground state of the one dimensional bond-charge Hubbard model. A paradigmatic quantum spin model in one dimension is the XYZ model, represented by the Hamiltonian 
\begin{eqnarray}
H&=&\frac{J}{4}\sum_{i=1}^N\left\{(1+g)\sigma_i^x\sigma_{i+1}^x+(1-g)\sigma_i^y\sigma_{i+1}^y+\Delta\sigma_i^z\sigma_{i+1}^z\right\}\nonumber \\
 &&+\frac{h_f}{2}\sum_{i=1}^N\sigma_i^z,
 \label{xyz}
\end{eqnarray}
where $J$ is the strength of the nearest-neighbour exchange interaction, $g$ is the $x-y$ anisotropy parameter, $\Delta$ and $h_f$ are the anisotropy, and the strength of the external magnetic field, respectively, in the $z$ direction. Several important one-dimensional quantum spin Hamiltonians emerge from Eq. (\ref{xyz}). For example, the Hamiltonian in Eq. (\ref{xyz}), with $g=0$ and $h_f=0$, represents the one-dimensional XXZ model, while for $\Delta=0$ and $g=1$, the model corresponds to the one-dimensional quantum Ising model in an external transverse magnetic field.  In Ref.~\cite{Qiu14}, the monogamy score of squared quantum discord is used to investigate the critical points of the one-dimensional XXZ model. See Fig. \ref{mono_qpt} for the variation of $\delta_{D^2}^\leftarrow$ against the anisotropy parameter, $\Delta$. Monogamy properties of other quantum correlation measures such as geometric discord \cite{Song13} and measurement induced disturbance \cite{Qin16} have also been investigated in the XXZ model. In an experimental study investigating the ground state of the one-dimensional quantum Ising model in a transverse magnetic field,  by using a nuclear magnetic resonance (NMR) setup, monogamy scores of negativity and quantum discord  are shown to distinguish between the cases of positive (frustrated phase) and negative (non-frustrated phase) values of $J$ in the ground state of the system \cite{Rao13}. It is interesting to note that monogamy of entanglement has been used to constrain the bipartite entanglement of resonating valence bond states \cite{Chandran07,Dhar11,Roy16}. 

\subsection{Quantum biological processes}

An interesting recent development in physics has been the investigation of the possibility of quantum effects in certain complex biological processes. In particular, light-harvesting protein complexes have been modeled to investigate photosynthetic processes in certain bacteria, with specific interest in the role of quantum coherence and quantum correlations  \cite{Engel07, Caruso09, Sarovar10, Lambert13}. Several studies have investigated the well-known Fenna-Mathews-Olson (FMO) complex, which mediates energy transfer from the receiving chromophores to the central reaction center, and attempted to characterize the efficiency of the energy transfer in terms of quantum correlations \cite{Caruso09, Sarovar10}. The role of monogamy of quantum correlations in the dynamics was recently investigated in Refs.~\cite{Zhu12,Chanda14}. In Ref.~\cite{Chanda14}, it is shown how the monogamy of quantum correlations, as quantified by negativity and quantum discord, is able to detect the arrangement of the different chromophore nodes in the FMO complex and support the predicted pathways for the transfer of excitation energy. The results also reiterate the predominance of multiparty quantum correlation measures over bipartite correlations between the nodes of the FMO complex.

\section{Conclusions}
\label{conclusion}

Like other no-go theorems \cite{Wootters82,Dieks82,Yuen86,Barnum96,Bell66,Kochen67,Pati00,Kalev08}  in quantum information science, in a multipartite domain, restrictions on sharability of quantum correlations, named as monogamy of quantum correlations, play a crucial role in achieving  successes in, and in understanding of several quantum information processing tasks. In this chapter, we have discussed the monogamy properties of  information-theoretic quantum correlations, specifically quantum discord, and highlighted their significant features. Computable multipartite quantum correlation measures are rare,  although there are a handful of bipartite quantum correlation measures, including quantum discord and several ``discord-like" measures, which are possible to calculate, at least numerically. The concept of monogamy opens up a new avenue where multipartite properties of a system can be studied via bipartite quantum correlations and becomes extremely useful to study different physical systems. This leads to another interesting aspect of  monogamy, namely, its application in several key phenomena in quantum physics, ranging from quantum communication to the emerging research on quantum spin models, quantum biology, and open quantum systems,  which is also reviewed.

\end{document}